\documentclass{aa}
\usepackage{epsfig}
\usepackage {graphicx}
\usepackage{amssymb}
\usepackage{amsmath}
\usepackage{multirow}
\usepackage {txfonts}
\usepackage{natbib}
\usepackage{pifont}

\begin{document}


\title{Analytical models of X-shape magnetic fields in galactic halos}
\author{Katia Ferri\`ere\inst{1} \and Philippe Terral\inst{1}}
\offprints{Katia Ferri\`ere}
\institute{$^{1}$ IRAP, Universit\'e de Toulouse, CNRS,
9 avenue du Colonel Roche, BP 44346, F-31028 Toulouse Cedex 4, France}
\date{Received  ; accepted }
\titlerunning{Models of X-shape magnetic fields}
\authorrunning{Katia Ferri\`ere}

\abstract
{External spiral galaxies seen edge-on exhibit X-shape magnetic fields 
in their halos.
Whether the halo of our own Galaxy also hosts an X-shape magnetic field
is still an open question.}
{We would like to provide the necessary analytical tools to test the hypothesis 
of an X-shape magnetic field in the Galactic halo.}
{We propose a general method to derive analytical models of divergence-free 
magnetic fields whose field lines are assigned a specific shape.
We then utilize our method to obtain four particular models of X-shape 
magnetic fields in galactic halos.
In passing, we also derive two particular models of predominantly horizontal
magnetic fields in galactic disks.
All our field models have spiraling field lines with spatially varying 
pitch angle.}
{Our four halo field models do indeed lead to X patterns in synthetic 
synchrotron polarization maps.
Their precise topologies can all be explained by the action of a wind
blowing outward from the galactic disk or from the galactic center.
In practice, our field models may be used for fitting purposes or as inputs 
to various theoretical problems.}
{}


\maketitle

\section{\label{intro}Introduction}

Low-frequency radio continuum observations provide a powerful tool to detect 
and measure magnetic fields in the disks and halos of external galaxies
\citep[see, e.g., the recent review by][]{beck&w_13}.
Polarization observations of nearby edge-on spiral galaxies show 
that most galactic disks harbor large-scale magnetic fields that are horizontal, 
i.e., parallel to the disk plane \citep[e.g.,][]{wielebinski&k_93, dumke&kwk_95}.
In the past few years, high-sensitivity polarization observations 
have also revealed the presence in galactic halos of magnetic fields 
forming a general X pattern.
These so-called X-shape magnetic fields are characterized by a vertical 
(i.e., perpendicular to the disk plane) component that increases with both 
galactic radius and height in the four quadrants
\citep{tullmann&dsu_00, soida_05, krause&wd_06, krause_09, heesen&kbd_09,
braun&hb_10, soida&kdu_11, haverkorn&h_12}.

At first thought, the denomination of X-shape magnetic fields
could be reminiscent of magnetic reconnection.
However, in contrast to X-type reconnection configurations, 
which truly form a complete X, the observed X-shape magnetic fields 
actually have the central part of the X missing --
in that sense, the term "X-shape" may be a little misleading.
More importantly, X-shape magnetic fields appear on global galactic scales,
which are vastly larger than the resistive scales on which reconnection 
presumably takes place.

Since X-shape magnetic fields are seen only at large distances from 
the galactic center, it is hard to guess what their complete magnetic 
topology is and how their field lines connect together, 
i.e., horizontally near the rotation axis or vertically across the disk plane.
This, in turn, makes it difficult to pin down the exact nature 
of X-shape magnetic fields and to trace back to their origin.
Various physical scenarios have been proposed in the literature,
involving either the operation of a classical galactic dynamo in the halo 
\citep{tullmann&dsu_00, soida_05}
or, more likely, the action of a large-scale galactic wind from the disk
\citep{tullmann&dsu_00, beck_08, heesen&kbd_09, soida&kdu_11}.
Alternatively, it has been suggested that the X pattern does not pertain
to the large-scale regular magnetic field,
but rather to a small-scale anisotropic random field
which would be associated with extremely elongated magnetic loops
produced by a spiky wind (Micha{\l} Hanasz, private communication).
Only rotation measure (RM) studies can distinguish between 
regular and anisotropic random fields.
In the cases of NGC~5775 \citep{soida&kdu_11} and NGC~4631 
(Mora \& Krause 2013), the large values and the smooth distributions
of the RMs measured toward their halos speak in favor of regular X-shape fields,
but little information is available for other galaxies.
In the present study, we will proceed on the premise that 
the X pattern is really an attribute of the large-scale regular field.

Now the question that naturally comes to mind is whether the halo 
of our own Galaxy also possesses an X-shape magnetic field.
The most widely used description of the Galactic halo field relies on 
the double-torus picture, originally sketched by \cite{han_02}
and later modeled by \cite{prouza&s_03, sun&rwe_08, jansson&f_12a},
where the halo field is purely azimuthal, forms a torus on each side 
of the Galactic plane and has opposite signs above and below the plane.
For the first time, \cite{jansson&f_12a,jansson&f_12b} added to 
this double-torus component a simple X-shape component, 
which they chose to be axisymmetric and purely poloidal.\footnote{
\label{foot_poloidal}
Throughout this paper, we will follow the usual convention of employing 
the term "poloidal" to refer to magnetic fields with only radial and 
vertical components, i.e., with no azimuthal component, 
in galactocentric cylindrical coordinates.
Strictly speaking, this terminology is correct only for axisymmetric
magnetic fields \citep[e.g.,][]{moffat_78}.
}
They found that this addition improved the overall fit 
to the RM and synchrotron data combined.

In this paper, we will not attempt to settle the question of 
whether the Galactic halo possesses or not an X-shape magnetic field,
but we will provide the analytical tools required to seriously address
this question in view of the existing observations.
More specifically, we will propose a general method to derive analytical 
models of divergence-free, X-shape magnetic fields that can be applied 
to galactic halos.
We will then derive four particular X-shape models having a small number 
of free parameters, and we will verify that these models truly produce 
X patterns in synchrotron polarization maps.

We emphasize that the method proposed here to obtain analytical field models
is very general, in the sense that it can be applied to all sorts of magnetic 
configurations, with or without an X shape.
Our four X-shape models, which represent a particular application 
of our general method to halo magnetic fields, have themselves 
broad applicability.
First of all, they can be applied to our own Galaxy,
with different purposes in mind.
For instance, they can be used to construct synthetic maps of RMs,
of synchrotron total and polarized intensities or of any other relevant 
observables, which can then be confronted to existing observational maps
with the aim of placing constraints on the properties of a putative 
X-shape magnetic field in the Galactic halo.
In a different perspective, our X-shape models can serve as a framework 
to study theoretical problems, such as cosmic-ray propagation 
and large-scale magnetohydrodynamic phenomena, in our Galaxy.
Similarly, they can be applied to external galaxies,
both to construct various synthetic maps which can help to constrain 
their true magnetic morphologies
and to study the same kind of theoretical problems as in our own Galaxy.

For pedagogical reasons, we will start by developing models 
of purely poloidal magnetic fields, 
and since our intended application is to galactic halos,
we will restrict our attention to poloidal fields with an overall X shape
(section~\ref{models_xshape}).
We will then extend our poloidal field models to three dimensions,
by including an azimuthal field component that is directly linked to
the poloidal component, as expected from dynamo theory 
(section~\ref{models_halo}).
Taking advantage of the analytical expressions obtained for magnetic fields 
in galactic halos, we will, with a little extra work, 
derive three-dimensional models of magnetic fields in galactic disks;
these disk field models can be utilized either independently 
or in combination with our halo field models (section~\ref{models_disk}).
As a necessary check of the relevance of our halo field models,
we will use them to simulate synchrotron polarization maps 
of a hypothetical external galaxy seen edge-on and make sure that these maps 
display the expected X patterns (section~\ref{polarization_maps}).
Finally, we will conclude our study with a short summary
(section~\ref{discussion}).

\section{\label{models_xshape}Analytical models of poloidal, 
X-shape magnetic fields}

\subsection{\label{models_xshape_euler}Euler potentials}

To model magnetic fields in galactic halos (in particular, in the halo 
of our own Galaxy), we will consider various magnetic configurations 
characterized by field lines that form a general X pattern.
For each considered configuration, we will adopt a simple analytical function 
to describe the shape of field lines, and we will derive the analytical 
expression of the corresponding magnetic field vector, ${\bf B}$, 
as a function of galactocentric cylindrical coordinates, $(r,\varphi,z)$.

A convenient way to proceed is to use the Euler potentials, 
$\alpha$ and $\beta$, defined such that 
\begin{equation}
\label{eq_euler}
{\bf B} = {\bf \nabla} \alpha \times {\bf \nabla} \beta 
\end{equation}
\citep{northrop_63, stern_66}.
A first great advantage of the Euler representation is that the magnetic field 
is automatically divergence-free.
Another important advantage, particularly relevant in the present context, 
is that field lines can be directly visualized.
Indeed, according to Eq.~(\ref{eq_euler}), any given field line 
is the line of intersection between a surface of constant $\alpha$ 
and a surface of constant $\beta$.
As an immediate consequence, each field line can be defined by, or labeled with,
a given pair $(\alpha,\beta)$.

In this section, we focus on the particular case when the magnetic field 
is poloidal ($B_\varphi = 0$), postponing discussion of the general 
three-dimensional case to section~\ref{models_halo}.
We can then identify one of the Euler potentials, say $\beta$, 
with galactic azimuthal angle:
\begin{equation}
\label{eq_beta}
\beta = \varphi \ ,
\end{equation}
whereupon Eq.~(\ref{eq_euler}) yields
\begin{eqnarray}
B_r & = & - \frac{1}{r} \ \left( \frac{\partial \alpha}{\partial z} \right)_r
\label{eq_Br_alpha} \\
B_\varphi & = & 0 
\label{eq_Bphi} \\
B_z & = & \frac{1}{r} \ \left( \frac{\partial \alpha}{\partial r} \right)_z \ ,
\label{eq_Bz_alpha}
\end{eqnarray}
where subscripts $r$ and $z$ indicate that the partial derivatives 
with respect to $z$ and $r$ are to be taken at constant $r$ and $z$, respectively.
It appears that $\alpha$ represents the (poloidal) magnetic flux
per unit azimuthal angle (taken with the appropriate sign)
between the considered field line $(\alpha,\varphi)$
and a reference field line $(\alpha_0,\varphi)$.

\subsection{\label{models_xshape_given_phi}Magnetic field in a given meridional plane} 

\begin{table*}
\caption{Brief description of our four models of poloidal, X-shape 
magnetic fields in galactic halos.}
\label{table_xshape}
\begin{minipage}[t]{2\columnwidth}
\centering
\renewcommand{\footnoterule}{}  
\begin{tabular}{c|ccc|cc|cc|cccc}
\hline
\hline
\noalign{\smallskip}
Model & 
Shape of &
\ \, Reference \footnote{
Prescribed radius or height at which field lines are labeled 
by their height or radius, respectively.
} &
\ \, Label of \footnote{
Height or radius of field lines at the reference coordinate.
} &
\ \, Reference \footnote{
Radial or vertical field at the reference coordinate
as a function of field line label.
} & 
\ \, Vertical \footnote{
For each model, the first line refers to the original parity
in section~\ref{models_xshape_given_phi},
and, when present, the second line refers to the reversed parity.
} & 
$B_r$ & $B_z$ &
\multicolumn{4}{c}
{Free parameters \footnote{
In general, the free parameters can vary with azimuthal angle, $\varphi$.
However, in the particular case considered in section~\ref{models_xshape_azim_depend},
only $B_1$ depends on $\varphi$ (through Eq.~\ref{eq_B1_cos}).
}\footnote{
$r_1$ is the reference radius in models~A and B 
and $|z_1|$ the modulus of the reference height in models~C and D;
$a$ is the parameter governing the opening of quadratic field lines
in models~A and C ($a > 0$);
$n$ is minus the power-law index of field lines asymptotic to
the $z$-axis in model~B ($n \ge 1$) 
or the $r$-axis in model~D ($n \ge \frac{1}{2}$);
$B_1$ is the peak value of the reference field, 
taken as the normalization field ($B_1 \gtrless 0$); 
$H$ is the exponential scale height (in models~A and B)
and $L$ the exponential scale length (in models~C and D) of the reference field
($H > 0$ and $L > 0$).
}
} \\
& field lines & coordinate & field lines & field & parity & & \\
\noalign{\smallskip}
\hline
\noalign{\smallskip}
\ \, A \footnote
{Model~A can only be used in non-axisymmetric magnetic configurations,
with the addition of an azimuthal field component near the $z$-axis,
to prevent the magnetic field from becoming singular at $r = 0$
(see section~\ref{models_xshape_azim_depend}).
} & 
Eq.~(\ref{eq_mfl_z_quadratic}) \ -- \ Fig.~\ref{figure_xshape}a & 
$r_1 > 0$ & $z_1$ \ (Eq.~\ref{eq_mfl_z1_quadratic}) & 
Eq.~(\ref{eq_B1_z1_even}) & sym & 
Eq.~(\ref{eq_Br_z1_quadratic}) & Eq.~(\ref{eq_Bz_z1_quadratic}) &
$r_1$ & $a$ & $B_1$ & $H$ \\
& & & & 
Eq.~(\ref{eq_B1_z1_odd}) & antisym & & & & & & \\
\noalign{\smallskip}
\hline
\noalign{\smallskip}
B & Eq.~(\ref{eq_mfl_z_divlinear}) \ -- \ Fig.~\ref{figure_xshape}b & 
$r_1 > 0$ & $z_1$ \ (Eq.~\ref{eq_mfl_z1_divlinear}) &
Eq.~(\ref{eq_B1_z1_even}) & sym & 
Eq.~(\ref{eq_Br_z1_divlinear}) & Eq.~(\ref{eq_Bz_z1_divlinear}) &
$r_1$ & $n$ & $B_1$ & $H$ \\
& & & & 
Eq.~(\ref{eq_B1_z1_odd}) & antisym & & & & & & \\
\noalign{\smallskip}
\hline
\noalign{\smallskip}
C & Eq.~(\ref{eq_mfl_r_quadratic}) \ -- \ Fig.~\ref{figure_xshape}c & 
$z_1 = 0$ & $r_1$ \ (Eq.~\ref{eq_mfl_r1_quadratic}) &
Eq.~(\ref{eq_B1_r1_even}) & antisym & 
Eq.~(\ref{eq_Br_r1_quadratic}) & Eq.~(\ref{eq_Bz_r1_quadratic}) &
& $a$ & $B_1$ & $L$ \\
\noalign{\smallskip}
\hline
\noalign{\smallskip}
D & Eq.~(\ref{eq_mfl_r_divlinear}) \ -- \ Fig.~\ref{figure_xshape}d &
$z_1 = |z_1| \ {\rm sign}~z$ & $r_1$ \ (Eq.~\ref{eq_mfl_r1_divlinear}) & 
Eq.~(\ref{eq_B1_r1_even}) & antisym & 
Eq.~(\ref{eq_Br_r1_divlinear}) & Eq.~(\ref{eq_Bz_r1_divlinear}) &
$|z_1|$ & $n$ & $B_1$ & $L$ \\
& & & & 
Eq.~(\ref{eq_B1_r1_odd}) & sym & & & & & & \\
\noalign{\smallskip}
\hline
\end{tabular}
\end{minipage}
\end{table*}

\begin{figure*}
\hspace*{-1cm}
\includegraphics[scale=0.65]{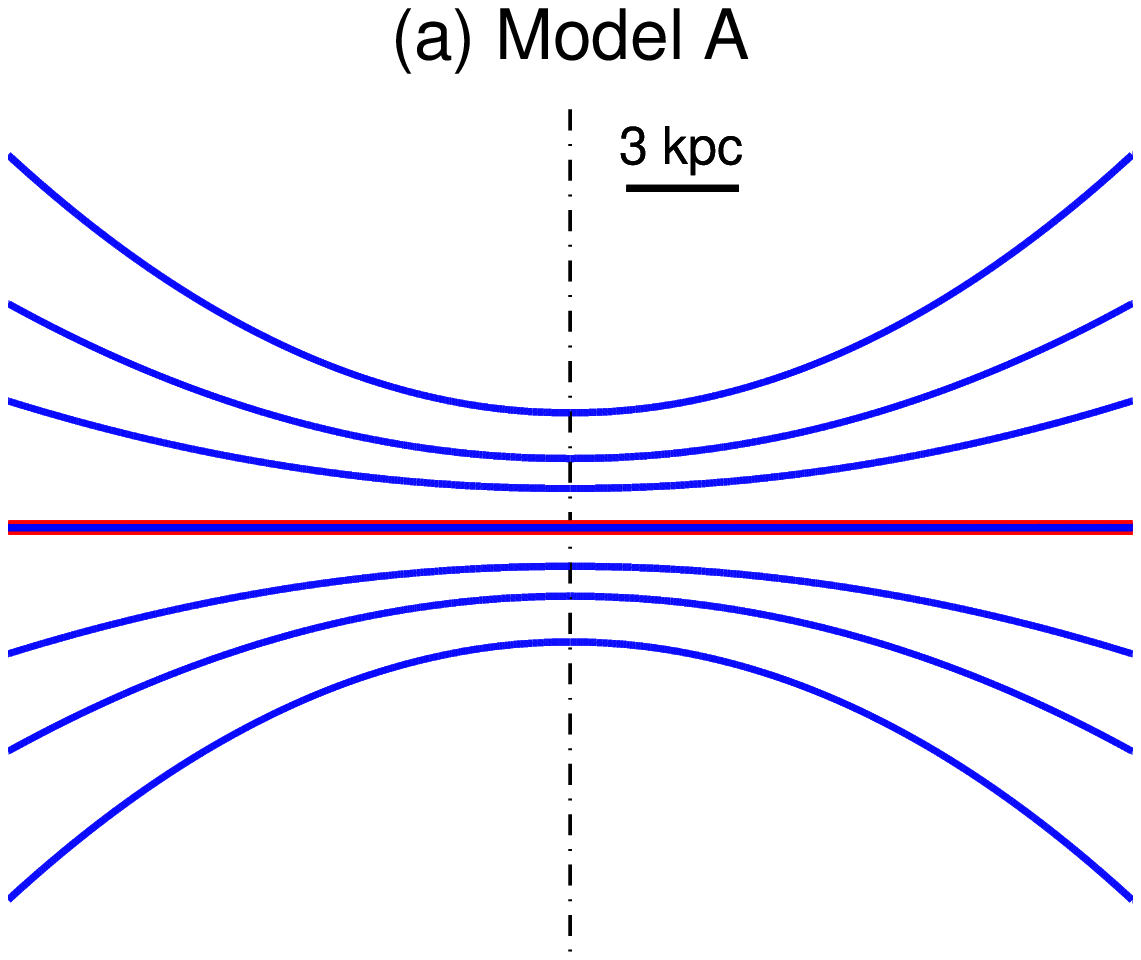}
\hspace*{1cm}
\includegraphics[scale=0.65]{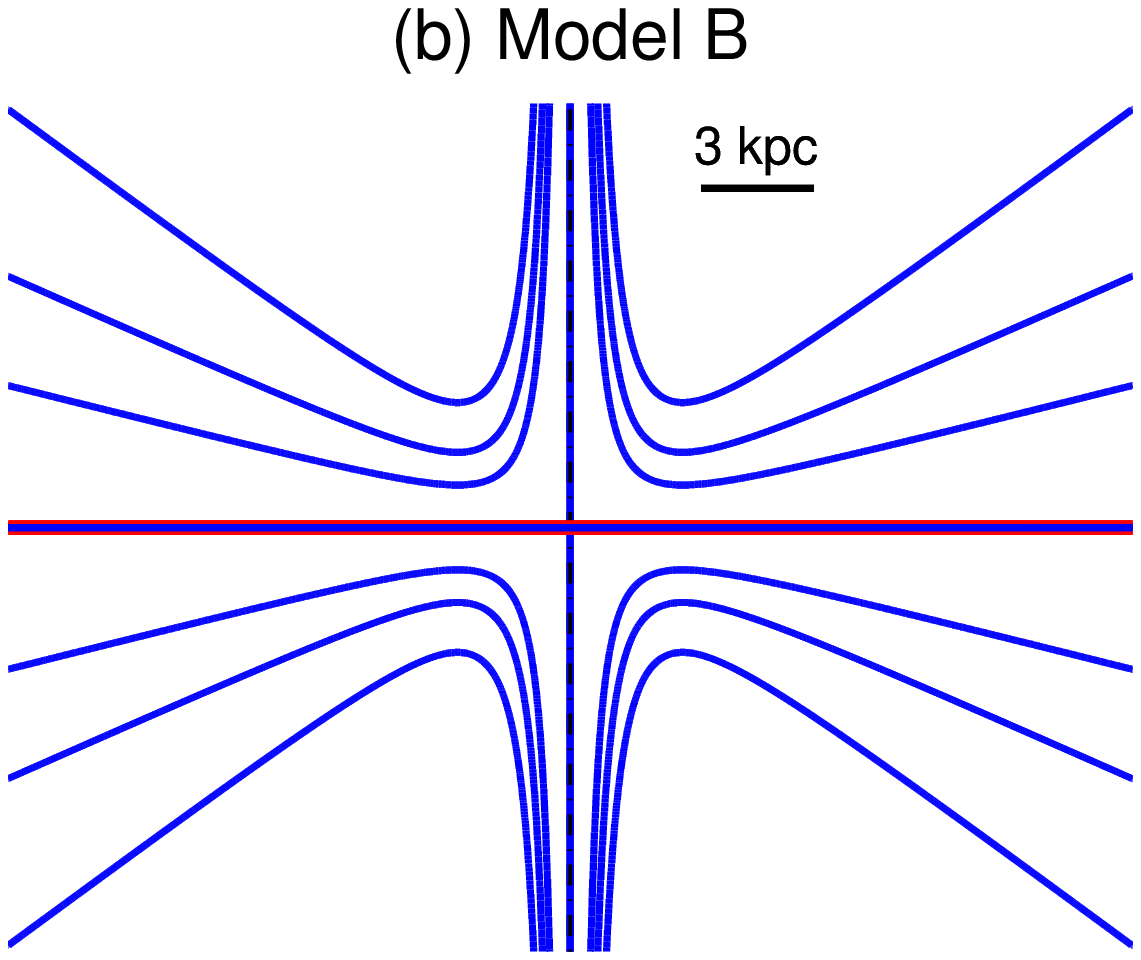}
\bigskip \\
\hspace*{-1cm}
\includegraphics[scale=0.65]{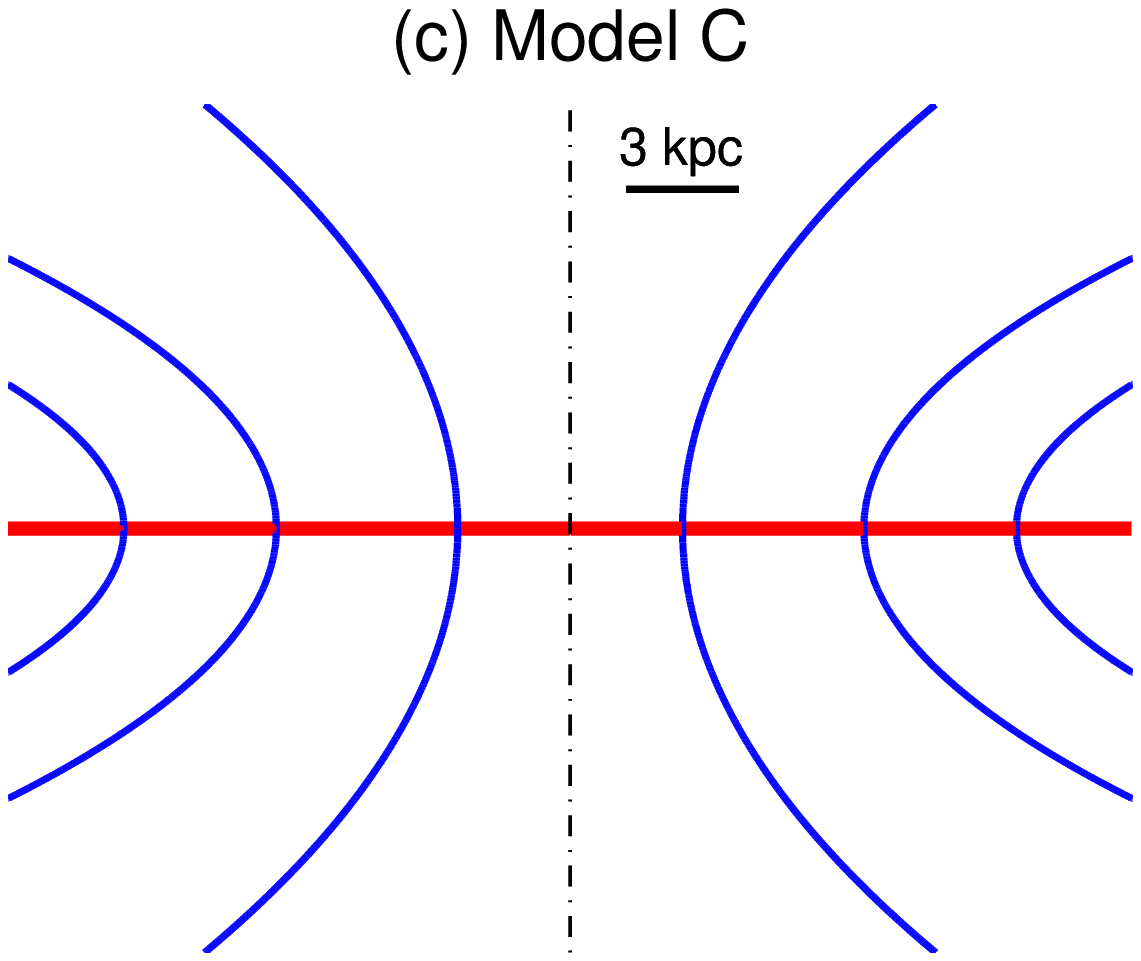}
\hspace*{1cm}
\includegraphics[scale=0.65]{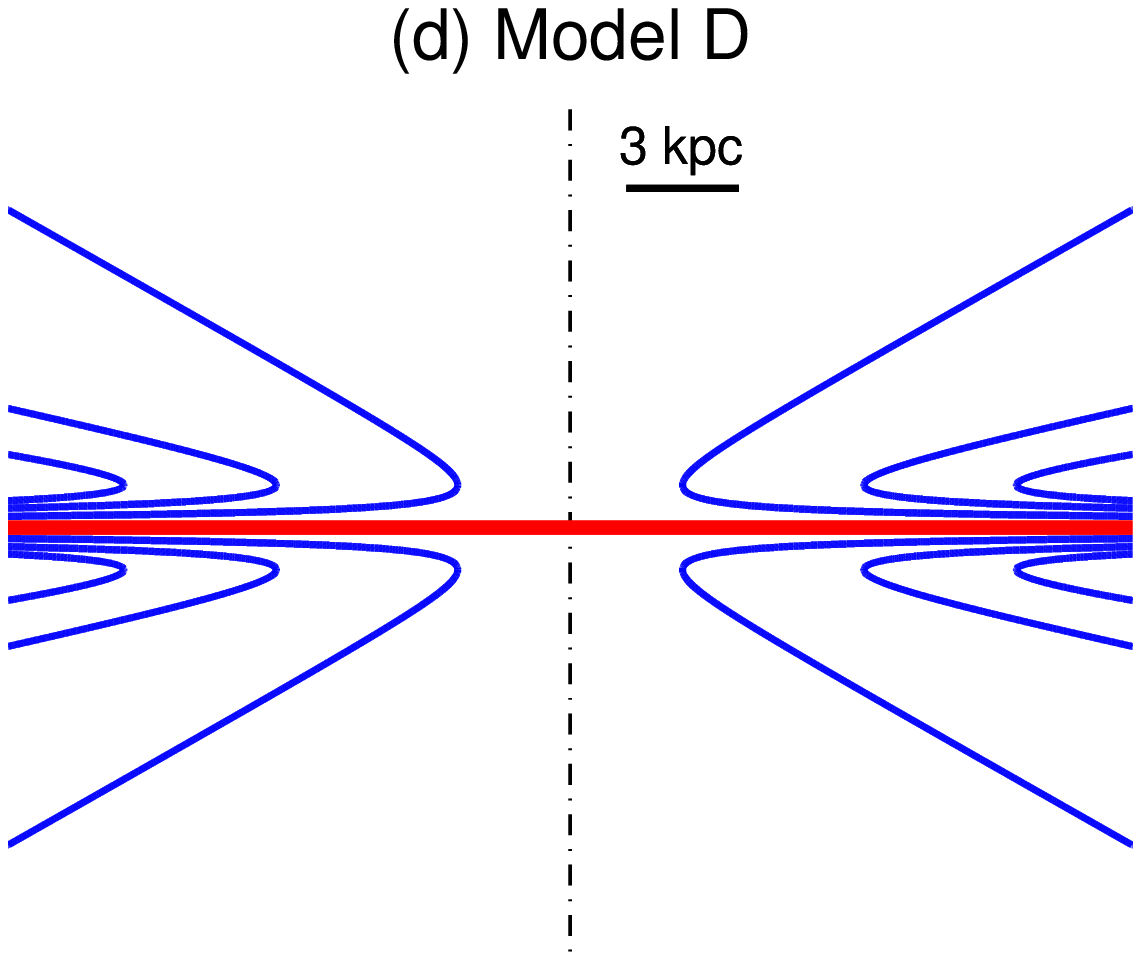}
\caption{Small set of field lines for each of our four models of poloidal, 
X-shape magnetic fields in galactic halos,
in a vertical plane through the galactic center:
(a) model~A (described by Eq.~\ref{eq_mfl_z_quadratic})
with $r_1 = 3$~kpc and $a = 1 / (10~{\rm kpc})^2$;
(b) model~B (Eq.~\ref{eq_mfl_z_divlinear})
with $r_1 = 3$~kpc and $n = 2$;
(c) model~C (Eq.~\ref{eq_mfl_r_quadratic})
with $a = 1 / (10~{\rm kpc})^2$ (and, as always, $z_1 = 0$);
and (d) model~D (Eq.~\ref{eq_mfl_r_divlinear})
with $|z_1| = 1.5$~kpc and $n = 2$.
Field lines are separated by a fixed magnetic flux per unit azimuthal angle;
thus, in models~A and B, their vertical spacing at $r_1$ is determined by 
Eq.~(\ref{eq_B1_z1_even}) or (\ref{eq_B1_z1_odd}) with $H = 1.5$~kpc,
while in models~C and D, their radial spacing at $z_1$ is determined by 
Eq.~(\ref{eq_B1_r1_even}) or (\ref{eq_B1_r1_odd}) with $L = 10$~kpc.
Each panel is $30~{\rm kpc} \times 30~{\rm kpc}$ in size
and is centered on the galactic center.
The trace of the galactic plane is indicated by the horizontal, red, solid line, 
and the $z$-axis by the vertical, black, dot-dashed line.
}
\label{figure_xshape}
\end{figure*}

For a poloidal, X-shape magnetic field, the equation of field lines
in a given meridional plane $\varphi$ can generally be written in the form
\begin{equation}
\label{eq_mfl_z_alpha}
z = F_z(\alpha,r)
\end{equation}
or
\begin{equation}
\label{eq_mfl_r_alpha}
r = F_r(\alpha,z) \ \cdot
\end{equation}
These two classes of field lines are successively studied 
in sections~\ref{models_xshape_mfl_z} and \ref{models_xshape_mfl_r}.

\subsubsection{\label{models_xshape_mfl_z}Field lines given by $z$ as a function of $r$}

Each field line can be identified by a value of $\alpha$ or, equivalently,
by a value of any monotonic function of $\alpha$.
A convenient such function is $z_1(\alpha)$, where $z_1$ denotes the height
of the considered field line at a specified (and fixed) radius $r_1$
(e.g., $r_1 = 3$~kpc).
The equation of field lines, Eq.~(\ref{eq_mfl_z_alpha}), can be rewritten 
in terms of $z_1$, as
\begin{equation}
\label{eq_mfl_z_z1}
z = f_z(z_1,r) \ ,
\end{equation}
and, accordingly, the expressions for the radial and vertical field components, 
Eqs.~(\ref{eq_Br_alpha}) and (\ref{eq_Bz_alpha}), can be transformed into
\begin{eqnarray}
B_r & = & - \frac{1}{r} \ \frac{d \alpha}{d z_1} 
\ \left( \frac{\partial z_1}{\partial z} \right)_r
\ = \ \frac{r_1}{r} \ B_r(r_1,z_1) 
\ \left( \frac{\partial z_1}{\partial z} \right)_r
\label{eq_Br_z1} \\
B_z & = & \frac{1}{r} \ \frac{d \alpha}{d z_1} 
\ \left( \frac{\partial z_1}{\partial r} \right)_z 
\ = \ - \frac{r_1}{r} \ B_r(r_1,z_1) 
\ \left( \frac{\partial z_1}{\partial r} \right)_z \ \cdot
\label{eq_Bz_z1}
\end{eqnarray}
Clearly, Eq.~(\ref{eq_Br_z1}) expresses magnetic flux conservation 
(in the radial direction) between $(r_1,z_1)$ and $(r,z)$,
while the ratio of Eq.~(\ref{eq_Bz_z1}) to Eq.~(\ref{eq_Br_z1}),
\begin{equation}
\label{eq_Bz_Br_z1}
\frac{B_z}{B_r} = \left( \frac{d z}{d r} \right)_{z_1} \ ,
\end{equation}
simply states that the magnetic field is tangent to field lines.

{\bf Model~A.}
To describe field lines with an X shape, we choose a function $z = f_z(z_1,r)$ 
such that $|z|$ increases as a positive power of $r$ for $r \to \infty$.
We discard the simple linear function of $r$, which leads to a cusp at $r = 0$,
and consider the quadratic function
\begin{equation}
\label{eq_mfl_z_quadratic}
z = z_1 \ \frac{1 + a \, r^2}{1 + a \, r_1^2} \ ,
\end{equation}
where $r_1$ and $a$ are two strictly positive free parameters
(see Figure~\ref{figure_xshape}a).
Let us emphasize that the values of $r_1$ and $a$ are unique and common 
to all field lines, in contrast to $z_1$, which has a different value 
on each field line.
By inverting Eq.~(\ref{eq_mfl_z_quadratic}) to obtain $z_1$ as a function of $(r,z)$:
\begin{equation}
\label{eq_mfl_z1_quadratic}
z_1 = z \ \frac{1 + a \, r_1^2}{1 + a \, r^2} \ ,
\end{equation}
calculating the partial derivatives of $z_1$ with respect to $z$ and $r$,
and introducing these derivatives into Eqs.~(\ref{eq_Br_z1}) and (\ref{eq_Bz_z1}),
we find for the radial and vertical field components:
\begin{eqnarray}
B_r & = & \frac{r_1}{r} \ \frac{z_1}{z} \ B_r(r_1,z_1)
\label{eq_Br_z1_quadratic} \\
B_z & = & \frac{2 \, a \, r_1 \, z_1}{1 + a \, r^2} \ B_r(r_1,z_1) \ \cdot
\label{eq_Bz_z1_quadratic}
\end{eqnarray}
To complete the model, it remains to specify the $z_1$-dependence of $B_r(r_1,z_1)$,
which, for simplicity, we choose to be linear-exponential:
\begin{equation} 
\label{eq_B1_z1_even}
B_r(r_1,z_1) = B_1 \ \frac{|z_1|}{H} \ \exp \left( - \frac{|z_1| - H}{H} \right) \ ,
\end{equation} 
with $B_1 \gtrless 0$ and $H > 0$ two additional free parameters.
As usual, the exponential factor ensures a fast enough decline at high $|z|$,
while the linear factor guarantees that the X-shape halo field remains weaker 
than the horizontal disk field (modeled in section~\ref{models_disk}) at low $|z|$.
The peak value of $B_r(r_1,z_1)$, reached at $|z_1| = H$, 
is $[B_r(r_1,z_1)]_{\rm peak} = B_1$.

It directly follows from Eq.~(\ref{eq_Br_z1_quadratic}) that 
$B_r \to \infty$ for $r \to 0$.
Evidently, this unphysical behavior arises because all field lines 
(from all meridional planes) converge to, or diverge from, the $z$-axis.
Nevertheless, in non-axisymmetric configurations (discussed in
section~\ref{models_xshape_azim_depend}),
a slight modification of the magnetic topology near the $z$-axis, 
involving the addition of a local azimuthal field component,
may be sufficient to remove the singularity.

{\bf Model~B.}
An alternative way of avoiding the singularity inherent in the parabolic form 
of Eq.~(\ref{eq_mfl_z_quadratic}) is to prevent field lines from reaching 
the $z$-axis.
This can be done by replacing the constant term in the right-hand side 
of Eq.~(\ref{eq_mfl_z_quadratic}) by a negative power of $r$.
The second term can remain quadratic, but it can also be replaced 
by a linear term without a cusp appearing at $r = 0$.
We prefer the second possibility, which we feel reproduces the observational
situation better.
We adjust the ratio of both terms by requiring that each field line reaches 
its minimum height at $(r_1,z_1)$.
Altogether, we write the equation of field lines as
\begin{equation}
\label{eq_mfl_z_divlinear}
z = \frac{1}{n+1} \ z_1 \ 
\left[ \left( \frac{r}{r_1} \right)^{-n} + n \, \frac{r}{r_1} \right] \ ,
\end{equation}
with $r_1$ and $n$ two strictly positive free parameters 
(see Figure~\ref{figure_xshape}b).
As before, we invert this equation in favor of $z_1$:
\begin{equation}
\label{eq_mfl_z1_divlinear}
z_1 = (n+1) \ z \ 
\left[ \left( \frac{r}{r_1} \right)^{-n} + n \, \frac{r}{r_1} \right]^{-1} \ ,
\end{equation}
and we insert the partial derivatives of $z_1$ into Eqs.~(\ref{eq_Br_z1}) 
and (\ref{eq_Bz_z1}), to obtain:
\begin{eqnarray}
B_r & = & \frac{r_1}{r} \ \frac{z_1}{z} \ B_r(r_1,z_1)
\label{eq_Br_z1_divlinear} \\
B_z & = & - \frac{n}{n+1} \ \frac{r_1 \, z_1^2}{r^2 \, z} \
\left[ \left( \frac{r}{r_1} \right)^{-n} - \frac{r}{r_1} \right] \ B_r(r_1,z_1) \ \cdot
\label{eq_Bz_z1_divlinear}
\end{eqnarray}
The $z_1$-dependence of $B_r(r_1,z_1)$ is again chosen to follow
Eq.~(\ref{eq_B1_z1_even}).
Finally, the requirement that $B_r$ and $B_z$ remain finite in the limit $r \to 0$,
where $B_r \propto r^{n-1} \ B_r(r_1,z_1) \propto r^{2n-1}$ 
and $B_z \propto r^{n-2} \ B_r(r_1,z_1) \propto r^{2n-2}$,
leads to the constraint $n \ge 1$.

In both model~A (Eqs.~(\ref{eq_Br_z1_quadratic}) -- (\ref{eq_Bz_z1_quadratic}))
and model~B (Eqs.~(\ref{eq_Br_z1_divlinear}) -- (\ref{eq_Bz_z1_divlinear})),
$B_r$ and $B_z$ are, respectively, even and odd functions of $z$.
The magnetic field is, therefore, symmetric with respect to the galactic 
midplane.\footnote{
\label{foot_sym}
Magnetic fields that are symmetric/antisymmetric with respect to the midplane 
are sometimes also referred to as quadrupolar/dipolar.
}
In principle, it is possible to obtain antisymmetric magnetic fields
with the same families of field lines, by choosing for $B_r(r_1,z_1)$ 
an odd function of $z_1$ -- for instance, by replacing the linear factor
in the right-hand side of Eq.~(\ref{eq_B1_z1_even}) by $(z_1/H)$:
\begin{equation} 
\label{eq_B1_z1_odd}
B_r(r_1,z_1) = B_1 \ \frac{z_1}{H} \ \exp \left( - \frac{|z_1| - H}{H} \right) \ \cdot
\end{equation}

\subsubsection{\label{models_xshape_mfl_r}Field lines given by $r$ as a function of $z$}

Here, each field line is identified by a value of the function $r_1(\alpha)$, 
where $r_1$ denotes the radius of the considered field line 
at a specified (and fixed) height $z_1$ (e.g., $z_1 = 0$).
In magnetic configurations where field lines do not cross the midplane,
one has to adopt two values of $z_1$ with opposite signs and select 
the positive/negative value of $z_1$ for field lines lying above/below 
the midplane. 
The equation of field lines, Eq.~(\ref{eq_mfl_r_alpha}), can be rewritten 
in terms of $r_1$, as
\begin{equation}
\label{eq_mfl_r_r1}
r = f_r(r_1,z) \ ,
\end{equation}
and the expressions for the radial and vertical field components, 
Eqs.~(\ref{eq_Br_alpha}) and (\ref{eq_Bz_alpha}), can be transformed into
\begin{eqnarray}
B_r & = & - \frac{1}{r} \ \frac{d \alpha}{d r_1} 
\ \left( \frac{\partial r_1}{\partial z} \right)_r
\ = \ - \frac{r_1}{r} \ B_z(r_1,z_1) 
\ \left( \frac{\partial r_1}{\partial z} \right)_r
\label{eq_Br_r1} \\
B_z & = & \frac{1}{r} \ \frac{d \alpha}{d r_1} 
\ \left( \frac{\partial r_1}{\partial r} \right)_z 
\ = \ \frac{r_1}{r} \ B_z(r_1,z_1) 
\ \left( \frac{\partial r_1}{\partial r} \right)_z \ \cdot
\label{eq_Bz_r1}
\end{eqnarray}
It is now Eq.~(\ref{eq_Bz_r1}) which expresses magnetic flux conservation
(in the vertical direction) between $(r_1,z_1)$ and $(r,z)$,
while the ratio of Eq.~(\ref{eq_Br_r1}) to Eq.~(\ref{eq_Bz_r1}),
\begin{equation}
\label{eq_Br_Bz_r1}
\frac{B_r}{B_z} = \left( \frac{d r}{d z} \right)_{r_1} \ ,
\end{equation}
corresponds again to the standard definition of field lines.

{\bf Model~C.}
For the same reasons as in section~\ref{models_xshape_mfl_z}, our first choice 
to describe X-shape field lines is the quadratic function
\begin{equation}
\label{eq_mfl_r_quadratic}
r = r_1 \ (1 + a \, z^2) \ ,
\end{equation}
obtained for $z_1 = 0$ and containing the free parameter $a > 0$ 
(see Figure~\ref{figure_xshape}c).
The inverse equation,
\begin{equation}
\label{eq_mfl_r1_quadratic}
r_1 = \frac{r}{1 + a \, z^2} \ ,
\end{equation}
combines with Eqs.~(\ref{eq_Br_r1}) and (\ref{eq_Bz_r1}) to yield
\begin{eqnarray}
B_r & = & \frac{2 \, a \, r_1^3 \, z}{r^2} \ B_z(r_1,z_1) 
\label{eq_Br_r1_quadratic} \\
B_z & = & \frac{r_1^2}{r^2} \ B_z(r_1,z_1) \ \cdot
\label{eq_Bz_r1_quadratic}
\end{eqnarray}
For the $r_1$-dependence of $B_z(r_1,z_1)$, we assume a simple exponential:
\begin{equation}
\label{eq_B1_r1_even}
B_z(r_1,z_1) = B_1 \ \exp \left( - \frac{r_1}{L} \right) \ ,
\end{equation}
with $B_1 \gtrless 0$ and $L > 0$ two free parameters.

It is interesting to note that, in contrast to the parabolic field 
derived in section~\ref{models_xshape_mfl_z} (Eqs.~(\ref{eq_Br_z1_quadratic}) --
(\ref{eq_Bz_z1_quadratic})), the parabolic field obtained here 
(Eqs.~(\ref{eq_Br_r1_quadratic}) -- (\ref{eq_Bz_r1_quadratic})) is perfectly regular.
The physical explanation is straightforward:
whereas the field lines governed by Eq.~(\ref{eq_mfl_z_quadratic}) 
all converge to, or diverge from, the same $z$-axis,
the field lines governed by Eq.~(\ref{eq_mfl_r_quadratic}) only cross
the $r$-axis of their own meridional plane.

{\bf Model~D.}
For completeness, we also examine the counterpart of Eq.~(\ref{eq_mfl_z_divlinear}),
\begin{equation}
\label{eq_mfl_r_divlinear}
r = \frac{1}{n+1} \ r_1 \ 
\left[ \left( \frac{z}{z_1} \right)^{-n} + n \, \frac{z}{z_1} \right] \ ,
\end{equation}
with $|z_1|$ and $n$ two strictly positive free parameters 
and $z_1$ required to take on the same sign as $z$,
i.e., $z_1 = |z_1| \ {\rm sign}~z$, such that the ratio $(z/z_1)$
is always positive (see Figure~\ref{figure_xshape}d).
With the inverse equation,
\begin{equation}
\label{eq_mfl_r1_divlinear}
r_1 = (n+1) \ r \ 
\left[ \left( \frac{z}{z_1} \right)^{-n} + n \, \frac{z}{z_1} \right]^{-1} \ ,
\end{equation}
Eqs.~(\ref{eq_Br_r1}) and (\ref{eq_Bz_r1}) reduce to
\begin{eqnarray}
B_r & = & - \frac{n}{n+1} \ \frac{r_1^3}{r^2 \, z} \
\left[ \left( \frac{z}{z_1} \right)^{-n} - \frac{z}{z_1} \right] \ B_z(r_1,z_1)
\label{eq_Br_r1_divlinear} \\
B_z & = & \frac{r_1^2}{r^2} \ B_z(r_1,z_1) \ \cdot
\label{eq_Bz_r1_divlinear}
\end{eqnarray}
The $r_1$-dependence of $B_z(r_1,z_1)$ is again assumed to follow 
Eq.~(\ref{eq_B1_r1_even}),
while the exponent $n$ must now satisfy the constraint $n \ge \frac{1}{2}$,
as can be seen from the behavior 
$B_r \propto z^{2n-1} \ B_z(r_1,z_1) \propto z^{2n-1}$ 
and $B_z \propto z^{2n} \ B_z(r_1,z_1) \propto z^{2n}$
in the limit $z \to 0$.

In both model~C (Eqs.~(\ref{eq_Br_r1_quadratic}) -- (\ref{eq_Bz_r1_quadratic}))
and model~D (Eqs.~(\ref{eq_Br_r1_divlinear}) -- (\ref{eq_Bz_r1_divlinear})),
$B_r$ and $B_z$ are, respectively, odd and even functions of $z$,
so that the magnetic field is anti-symmetric with respect to the midplane
(see footnote~\ref{foot_sym}).
Model~D can also be made symmetric, by linking the sign of $B_z(r_1,z_1)$
to the sign of $z_1$ -- for instance, by multiplying the right-hand side 
of Eq.~(\ref{eq_B1_r1_even}) by ${\rm sign}~z_1$: 
\begin{equation}
\label{eq_B1_r1_odd}
B_z(r_1,z_1) = B_1 \ {\rm sign}~z_1 \ \exp \left( - \frac{r_1}{L} \right) \ \cdot
\end{equation}
On the other hand, model~C cannot be made symmetric, because field lines
actually cross the midplane, so that $B_z$ cannot change sign across the midplane.

\subsection{\label{models_xshape_vert_sym}Vertical symmetry}

In section~\ref{models_xshape_given_phi}, we presented four different 
analytical models of poloidal, X-shape magnetic fields for application
to galactic halos.
Our models in their original form are either symmetric or antisymmetric 
with respect to the midplane, 
but their parity can generally be reversed, simply by flipping the parity 
of the function chosen for $B_r(r_1,z_1)$ or $B_z(r_1,z_1)$.
Thus, models~A and B are originally symmetric, with $B_r(r_1,z_1)$ 
given by an even function of $z_1$ (Eq.~\ref{eq_B1_z1_even}),
but they also have an antisymmetric version, obtained by turning $B_r(r_1,z_1)$ 
into an odd function of $z_1$ (Eq.~\ref{eq_B1_z1_odd}).
Similarly, models~C and D are originally antisymmetric, with $B_z(r_1,z_1)$  
independent of the sign of $z_1$ (Eq.~\ref{eq_B1_r1_even}),
but model~D also has a symmetric version, obtained by linking the sign 
of $B_z(r_1,z_1)$ to the sign of $z_1$ (Eq.~\ref{eq_B1_r1_odd}).
For convenience, the descriptive equations and parameters of the four models, 
in their original and modified versions,
are summarized in Table~\ref{table_xshape}.

Let us now discuss the possible physical origin of X-shape magnetic fields,
in connection with their vertical parity.
The natural tendency of a conventional galactic dynamo is
to generate a symmetric, mainly horizontal, field in the disk 
and an antisymmetric, possibly dipole-like, field in the halo
\citep[e.g.,][]{sokoloff&s_90}.
In general, though, the disk and halo fields are coupled together
and do not evolve independently.
What usually happens is that one field enslaves and imposes its natural parity 
to the other, so that the total (disk + halo) field ends up being
either symmetric (if the disk dynamo dominates) 
or antisymmetric (if the halo dynamo dominates) \citep{moss&s_08}.
However, in the presence of a moderate galactic wind, the total field 
can sometimes have mixed parity, being approximately symmetric
in the disk and approximately antisymmetric in the halo \citep{moss&sbk_10}.

All-sky RM maps point to such a mixed-parity magnetic configuration 
in our Galaxy \citep{rand&l_94, oren&w_95, han&mbb_97, han&mq_99, frick&sss_01}.
The situation in external galaxies remains uncertain,
but the available data would also tend to suggest mixed parity.
In the particular case of NGC~5775, a preliminary derivation of RMs 
toward the halo revealed a flip in the RM sign across the midplane, 
consistent with an antisymmetric poloidal halo field \citep{tullmann&dsu_00};
however, a subsequent RM analysis using additional data 
at a shorter wavelength (almost free of Faraday rotation)
found the RM sign to be continuous across, and out to large distances from,
the midplane \citep{soida&kdu_11}.
In a more comprehensive study, RM synthesis of a sample 
of 21 WSRT-SINGS galaxies suggests that the halo field generally possesses 
a symmetric (axisymmetric spiral quadrupole) component,
which would be a mere extension of the symmetric thick-disk field, 
plus an antisymmetric (almost purely poloidal dipole) component, 
which would become dominant in the upper halo \citep{braun&hb_10}.

Regardless of vertical parity, conventional dynamo fields do not 
naturally come with an X shape, but they can potentially acquire one 
under the action of a galactic wind 
\citep[e.g.,][]{brandenburg&dms_93, moss&sbk_10}.
Quite naturally, symmetric X-shape fields would be related to the disk dynamo,
while antisymmetric X-shape fields would be related to the halo dynamo.
More specifically, the former could be explained with a wind originating 
near the galactic plane and advecting the disk dynamo field into the halo,
while the latter could be explained with a wind blowing from the base 
of the halo and stretching out the halo dynamo field into an X shape.

From a morphological point of view, in models~A, B and D, where low-$|z|$ 
field lines tend to follow the galactic plane, the X-shape field could 
conceivably be traced back to an initially horizontal disk field,
assuming the wind has the following properties:
in model~A, the wind would blow obliquely outward from the disk
(possibly due to pressure gradients; see the hydrodynamical simulations 
of \cite{dallavecchia&s_08});
in model~B, the wind would have a weak oblique component emanating from the disk
plus a strong bipolar-jet component emanating from the galactic center 
(most likely driven by an AGN or by a nuclear starburst);
and in model~D, the wind would take the form of a champagne flow 
originating in the central region (probably due to strong stellar activity).
In these three models, the X-shape field could {\it a priori} be either 
symmetric or antisymmetric, but if it is indeed connected to the disk field 
and if the latter has evolved to and retained its natural parity
(two non-trivial conditions), the X-shape field should be symmetric.
In contrast, in model~C, where field lines cross the galactic plane,
the X-shape field is easily reconciled with an initially dipole-like halo field,
which was blown open either along the plane by an oblique wind from the disk 
or along the $z$-axis by a champagne flow from the central region.
In this model, the X-shape field would obvioulsy keep the antisymmetric character 
of the initial dipole-like field.

Somewhat different conclusions on the vertical parity of the halo field
emerged from the galactic dynamo calculations of \cite{moss&sbk_10}.
They found that halo fields that are antisymmetric tend to come with an X shape
-- a tendency which is reinforced by the presence of a modest wind,
whereas halo fields that have become symmetric, presumably under the influence
of the disk, tend to lose all resemblance to an X pattern.

\subsection{\label{models_xshape_azim_depend}Azimuthal variation}

The expressions derived in section~\ref{models_xshape_given_phi}
and discussed in section~\ref{models_xshape_vert_sym}
are valid in any given meridional plane.
They can be generalized to full three-dimensional space by letting 
the free parameters (listed in the last column of Table~\ref{table_xshape})
be functions of the azimuthal angle, $\varphi$.

Here, we consider the very simple case when the only free parameter
allowed to vary with $\varphi$ is the normalization field, $B_1$, 
defined as $B_1 = B_r(r_1,H)$ in models~A and B (see Eqs.~(\ref{eq_B1_z1_even})
and (\ref{eq_B1_z1_odd}))
and $B_1 = B_z(0,|z_1|)$ in models~C and D (see Eqs.~(\ref{eq_B1_r1_even})
and (\ref{eq_B1_r1_odd})).
We further assume that the variation of $B_1$ with $\varphi$ is sinusoidal:
\begin{equation}
\label{eq_B1_cos}
B_1(\varphi) = B_{1\star} \ \cos \big( m \, (\varphi - \varphi_\star) \big) \ ,
\end{equation}
with $m$ the azimuthal wavenumber (taken to be positive)
and $B_{1\star}$ and $\varphi_\star$ two free parameters (allowed to take on
either sign).
Clearly, $m = 0$ corresponds to a purely axisymmetric magnetic field,
$m = 1$ to a bisymmetric field, $m = 2$ to a quadrisymmetric field, 
and so on.

We now return to model~A and show that the nature of its singularity at $r = 0$ 
depends on whether the magnetic field is axisymmetric or not.
Consider a cylinder of radius $R$ and semi-infinite length, 
centered on the $z$-axis and extending from $z = 0$ to $z \to +\infty$.
The magnetic flux across the surface of this cylinder reads
\begin{eqnarray}
\Phi_{\rm m} & = & 
R \ \int_0^{2\pi} d\varphi \ \int_{0}^{+\infty} dz \ B_r(R,\varphi,z)
\nonumber \\
& - & \int_0^R dr \ r \ \int_0^{2\pi} d\varphi \ B_z(r,\varphi,0)
\nonumber \\
& + & \lim_{Z \to {+\infty}} \int_0^R dr \ r \ \int_0^{2\pi} d\varphi \ 
B_z(r,\varphi,Z) \ ,
\label{eq_flux_cyl}
\end{eqnarray}
where the three terms in the right-hand side represent, respectively,
the flux through the side surface at $r = R$, 
the flux through the bottom surface at $z = 0$
and the flux through the top surface at $z \to +\infty$.
These terms can be calculated using Eqs.~(\ref{eq_Br_z1_quadratic}) 
and (\ref{eq_Bz_z1_quadratic}), in conjunction with Eq.~(\ref{eq_mfl_z1_quadratic})
and Eq.~(\ref{eq_B1_z1_even}) or (\ref{eq_B1_z1_odd}).
The second and third terms are easily shown to vanish, leaving only
\begin{equation}
\label{eq_flux_cyl_side}
\Phi_{\rm m} = (\exp{1}) \ H \ r_1 \ \int_0^{2\pi} d\varphi \ B_1(\varphi) \ \cdot
\end{equation}
Now, the divergence-free condition implies that $\Phi_{\rm m}$ must vanish,
which, in turn, implies 
\begin{equation}
\label{eq_flux_zero}
\int_0^{2\pi} d\varphi \ B_1(\varphi) = 0 \ \cdot
\end{equation}
For the sinusoidally-varying field given by Eq.~(\ref{eq_B1_cos}), 
this is possible only if the azimuthal wavenumber $m \not= 0$.

In consequence, an axisymmetric ($m = 0$) magnetic field in model~A
is intrinsically unphysical, in the sense that it automatically leads 
to a non-vanishing magnetic flux out of, or into, the considered cylinder
-- in contradiction with the divergence-free condition.
This occurs because, in the axisymmetric case, the magnetic field
on the side surface of the cylinder points everywhere outward or inward.
It then follows that the magnetic field must necessarily become singular 
along the axis of the cylinder.

In contrast, a non-axisymmetric ($m \not= 0$) magnetic field has no net
magnetic flux across the side surface of the cylinder, because the magnetic field
points alternatively outward and inward.
The singularity found at $r = 0$ arises only because the magnetic field
is required to be poloidal everywhere and to become radial as $r \to 0$, 
so that field lines have no other choice but to run into each other at $r = 0$.
This unacceptable situation could be avoided by simply introducing an azimuthal 
field component within a small radius, which would enable field lines
to remain separate from each other,
except for connecting two by two to ensure field line continuity.
In this manner, the magnetic field would remain finite everywhere.

Hence, model~A can be retained to describe non-axisymmetric magnetic fields
outside a small radius.
Inside this radius, an azimuthal component must be added to the poloidal
field so as to remove the singularity at $r = 0$.
Note that, in practice, it may not be necessary to specify the exact form 
of the azimuthal component; knowing that the model can be made physical 
may be sufficient.

Model~B does not become singular at $r = 0$, even in the axisymmetric case, 
because all the magnetic flux entering/leaving through the side surface 
of the cylinder finds its way out/in through the top surface 
(see Figure~\ref{figure_xshape}b).
This can be verified mathematically by examining the three terms 
in the right-hand side of Eq.~(\ref{eq_flux_cyl}).
The first term (flux through the side surface)
has the same expression as in model~A,
because in both models $z_1$ is a linear function of $z$
(see Eq.~(\ref{eq_mfl_z1_quadratic}) for model~A
and Eq.~(\ref{eq_mfl_z1_divlinear}) for model~B).
The second term (flux through the bottom surface) vanishes,
because, like in model~A, $B_z = 0$ at $z = 0$
(see Eq.~(\ref{eq_Bz_z1_quadratic}) for model~A
and Eq.~(\ref{eq_Bz_z1_divlinear}) for model~B).
The difference between the two models comes from 
the third term (flux through the top surface), which vanishes in model~A, 
but is equal and opposite to the first term (flux through the side surface)
in model~B.

\section{\label{models_halo}Analytical models of three-dimensional magnetic fields
in galactic halos}

\subsection{\label{models_halo_spiral}Magnetic field with a spiral horizontal
component}

In this section, we let the magnetic field have both poloidal and azimuthal 
components.
Moreover, based on observations of external face-on spiral galaxies, 
we assume that the magnetic field projected onto the galactic plane 
forms a spiral pattern,
in which each spiral line satisfies an equation of the type
\begin{equation}
\label{eq_mfl_spiral}
\varphi = \varphi_0 + f_\varphi (r) \ ,
\end{equation}
where $\varphi_0$ is the azimuthal angle at which the line passes through 
a chosen (and fixed) radius $r_0$,
and $f_\varphi (r)$ is a monotonic function of $r$ which vanishes at $r_0$.
The pitch angle of the spiral, $p$, defined as the angle between the tangent 
to the spiral and the azimuthal direction, is given by
\begin{equation}
\label{eq_pitchangle}
\cot p = \frac{r \ d\varphi}{dr}
= r \ \frac{df_\varphi}{dr} \ \cdot
\end{equation}
In most studies of galactic magnetic fields, $p$ is supposed to be constant,
which implies a logarithmic spiral, with
\begin{equation}
\label{eq_log_spiral}
f_\varphi (r) = \cot p \ \ln \frac{r}{r_0} \ \cdot
\end{equation}
However, the assumption of constant pitch angle is generally not realistic,
either in external galaxies or in our own Galaxy
\citep[e.g.,][]{fletcher_10}.
This unrealistic assumption is not made in our formalism, which naturally 
allows the pitch angle to vary with galactic radius.

When the magnetic field has an azimuthal component ($B_\varphi \not= 0$),
$\varphi$ is no longer constant along field lines and, therefore, may 
no longer serve as one of the Euler potentials.
Instead, one can use $\varphi_0$ and, accordingly, replace Eq.~(\ref{eq_beta}) by
\begin{equation}
\label{eq_beta_spiral}
\beta = \varphi_0 = \varphi - f_\varphi (r) \ \cdot
\end{equation}
With this choice of $\beta$, Eq.~(\ref{eq_euler}) gives for the radial,
azimuthal and vertical field components
\begin{eqnarray}
B_r & = & - \frac{1}{r} \ 
\left( \frac{\partial \alpha}{\partial z} \right)_{r,\varphi}
\nonumber \\ 
B_\varphi & = & - \frac{df_\varphi}{dr} \ 
\left( \frac{\partial \alpha}{\partial z} \right)_{r,\varphi}
\nonumber \\ 
B_z & = & \frac{1}{r} \ 
\left( \frac{\partial \alpha}{\partial r} \right)_{\varphi,z}
+ \frac{1}{r} \ \frac{df_\varphi}{dr} \ 
\left( \frac{\partial \alpha}{\partial \varphi} \right)_{r,z} 
\nonumber 
\end{eqnarray}
or, equivalently,
\begin{eqnarray}
B_r & = & - \frac{1}{r} \ 
\left( \frac{\partial \alpha}{\partial z} \right)_{r,\varphi_0}
\label{eq_Br_alpha_spiral} \\
B_\varphi & = & r \ \frac{df_\varphi}{dr} \ B_r
\ = \ \cot p \ B_r
\label{eq_Bphi_alpha_spiral} \\
B_z & = & \frac{1}{r} \ 
\left( \frac{\partial \alpha}{\partial r} \right)_{\varphi_0,z}
\ \cdot
\label{eq_Bz_alpha_spiral}
\end{eqnarray}
Eqs.~(\ref{eq_Br_alpha_spiral}) and (\ref{eq_Bz_alpha_spiral}) 
are the direct counterparts of Eqs.~(\ref{eq_Br_alpha}) and (\ref{eq_Bz_alpha})
in section~\ref{models_xshape};
the only difference between both pairs of equations is that the partial derivatives 
of $\alpha$ in Eqs.~(\ref{eq_Br_alpha}) and (\ref{eq_Bz_alpha})
are taken at constant $\varphi$, i.e., in meridional planes, 
whereas those in Eqs.~(\ref{eq_Br_alpha_spiral}) and (\ref{eq_Bz_alpha_spiral}) 
are taken at constant $\varphi_0$, i.e., along the spiral surfaces defined 
by Eq.~(\ref{eq_mfl_spiral}).
Eq.~(\ref{eq_Bphi_alpha_spiral}), for its part, states that the horizontal 
magnetic field has pitch angle $p$ and, therefore, follows the spiral lines 
defined in the galactic plane by Eq.~(\ref{eq_mfl_spiral}).
Let us emphasize again that $p$ is not required to remain constant along field lines.

\subsection{\label{models_halo_full}Magnetic field with a spiral horizontal 
component and an X-shape poloidal component}

The similarity between the pair of Eqs.~(\ref{eq_Br_alpha}) -- (\ref{eq_Bz_alpha})
and the pair of Eqs.~(\ref{eq_Br_alpha_spiral}) -- (\ref{eq_Bz_alpha_spiral})
indicates that the derivation presented in section~\ref{models_xshape_given_phi}
can be repeated here, with this only difference that the working space is now
a spiral surface of constant $\varphi_0$ rather than a meridional plane 
of constant $\varphi$.
Likewise, the azimuthal variation introduced 
in section~\ref{models_xshape_azim_depend} can be taken up here,
with Eq.~(\ref{eq_B1_cos}) written as a function of $\varphi_0$
instead of $\varphi$.
For future reference, we recall that 
\begin{equation}
\label{eq_phi0}
\varphi_0 = \varphi - f_\varphi (r) 
\end{equation}
(see Eq.~\ref{eq_mfl_spiral}).

In the same spirit as in section~\ref{models_xshape_given_phi}, we distinguish 
between two classes of poloidal field lines, which we successively study
in sections~\ref{models_halo_mfl_z} and \ref{models_halo_mfl_r}.
Note that we only need to worry about the radial and vertical field components, 
as the azimuthal field, given by Eq.~(\ref{eq_Bphi_alpha_spiral}),
can easily be obtained from the radial field once the shape of the spiral
has been specified.

\subsubsection{\label{models_halo_mfl_z}Poloidal field lines given by $z$ 
as a function of $r$}

As in section~\ref{models_xshape_mfl_z}, we start by adopting a reference 
radius $r_1$.
Each field line can then be identified by the coordinates $(\varphi_1,z_1)$
at which it intersects the cylinder of radius $r_1$.
Since the reference radius, $r_1$, 
and the normalization radius of the spiral, $r_0$,
both have unique values (common to all field lines), 
we may, for convenience, let $r_1 = r_0$, and hence $\varphi_1 = \varphi_0$.

Following the procedure of section~\ref{models_xshape_mfl_z},
on a surface of constant $\varphi_0$ rather than constant $\varphi$, 
we transform Eqs.~(\ref{eq_Br_alpha_spiral}) and (\ref{eq_Bz_alpha_spiral}) into
\begin{eqnarray}
B_r & = & \frac{r_1}{r} \ B_r(r_1,\varphi_0,z_1) 
\ \left( \frac{\partial z_1}{\partial z} \right)_r
\label{eq_Br_z1_spiral} \\
B_z & = & - \frac{r_1}{r} \ B_r(r_1,\varphi_0,z_1) 
\ \left( \frac{\partial z_1}{\partial r} \right)_z \ ,
\label{eq_Bz_z1_spiral}
\end{eqnarray}
with $r_1$ the reference radius,
$z_1$ set by our adopted X shape (Eq.~(\ref{eq_mfl_z1_quadratic}) in model~A
and Eq.~(\ref{eq_mfl_z1_divlinear}) in model~B), 
and $\varphi_0$ set by our adopted spiral (Eq.~\ref{eq_phi0}).
A comparison with Eqs.~(\ref{eq_Br_z1}) -- (\ref{eq_Bz_z1})
shows that the expressions of $B_r$ and $B_z$
obtained in section~\ref{models_xshape_mfl_z} remain valid here
provided only that $B_r(r_1,z_1)$ be replaced by $B_r(r_1,\varphi_0,z_1)$.

Thus, in {\bf model~A}, 
Eqs.~(\ref{eq_Br_z1_quadratic}) -- (\ref{eq_Bz_z1_quadratic}) become
\begin{eqnarray}
B_r & = & \frac{r_1}{r} \ \frac{z_1}{z} \ B_r(r_1,\varphi_0,z_1)
\label{eq_Br_z1_quadratic_spiral} \\
B_z & = & \frac{2 \, a \, r_1 \, z_1}{1 + a \, r^2} \ B_r(r_1,\varphi_0,z_1) \ ,
\label{eq_Bz_z1_quadratic_spiral}
\end{eqnarray}
and in {\bf model~B},
Eqs.~(\ref{eq_Br_z1_divlinear}) -- (\ref{eq_Bz_z1_divlinear}) become
\begin{eqnarray}
B_r & = & \frac{r_1}{r} \ \frac{z_1}{z} \ B_r(r_1,\varphi_0,z_1)
\label{eq_Br_z1_divlinear_spiral} \\
B_z & = & - \frac{n}{n+1} \ \frac{r_1 \, z_1^2}{r^2 \, z} \
\left[ \left( \frac{r}{r_1} \right)^{-n} - \frac{r}{r_1} \right] \
B_r(r_1,\varphi_0,z_1) \ \cdot
\label{eq_Bz_z1_divlinear_spiral}
\end{eqnarray}
We assume that $B_r(r_1,\varphi_0,z_1)$ has the same vertical variation 
as in section~\ref{models_xshape_mfl_z} (see Eqs.~(\ref{eq_B1_z1_even}) and
(\ref{eq_B1_z1_odd})) and that its azimuthal variation is governed by 
Eq.~(\ref{eq_B1_cos}) written as a function of $\varphi_0$ instead of $\varphi$ 
(see first paragraph of section~\ref{models_halo_full}).
The result is 
\begin{equation}
\label{eq_B1_z1_spiral}
B_r(r_1,\varphi_0,z_1) = B_r(r_1,z_1) \ 
\cos \big( m \, (\varphi_0 - \varphi_\star) \big) \ ,
\end{equation}
with $B_r(r_1,z_1)$ given by Eq.~(\ref{eq_B1_z1_even}) for symmetric magnetic fields 
and by Eq.~(\ref{eq_B1_z1_odd}) for antisymmetric magnetic fields,
$m$ the azimuthal wavenumber and $\varphi_\star$ a free parameter.

\subsubsection{\label{models_halo_mfl_r}Poloidal field lines given by $r$ 
as a function of $z$}

Here, we adopt a reference height $z_1$ -- or two opposite-signed values of $z_1$ 
if field lines do not cross the midplane.
Each field line can then be identified by the coordinates $(r_1,\varphi_1)$
at which it intersects the horizontal plane -- or one of the two horizontal planes
-- of height $z_1$.
Since the value of $r_1$ now depends on the considered field line,
we may no longer equate $r_1$ to $r_0$, which must remain a separate 
free parameter.
According to Eq.~(\ref{eq_mfl_spiral}), the azimuthal angles associated 
with $r_1$ and $r_0$ are related through 
$\varphi_1 = \varphi_0 + f_\varphi (r_1)$.

In the same fashion as in section~\ref{models_xshape_mfl_r},
we transform Eqs.~(\ref{eq_Br_alpha_spiral}) and (\ref{eq_Bz_alpha_spiral}) into
\begin{eqnarray}
B_r & = & - \frac{r_1}{r} \ B_z(r_1,\varphi_0,z_1) 
\ \left( \frac{\partial r_1}{\partial z} \right)_r
\label{eq_Br_r1_spiral} \\
B_z & = & \frac{r_1}{r} \ B_z(r_1,\varphi_0,z_1) 
\ \left( \frac{\partial r_1}{\partial r} \right)_z \ ,
\label{eq_Bz_r1_spiral}
\end{eqnarray}
with $z_1$ the reference height,
$r_1$ set by our adopted X shape (Eq.~(\ref{eq_mfl_r1_quadratic}) in model~C
and Eq.~(\ref{eq_mfl_r1_divlinear}) in model~D),
and $\varphi_0$ set by our adopted spiral (Eq.~\ref{eq_phi0}).
A comparison with Eqs.~(\ref{eq_Br_r1}) -- (\ref{eq_Bz_r1})
shows that the results obtained in section~\ref{models_xshape_mfl_r} 
remain valid with $B_z(r_1,z_1)$ replaced by $B_z(r_1,\varphi_0,z_1)$.

Thus, in {\bf model~C},
Eqs.~(\ref{eq_Br_r1_quadratic}) -- (\ref{eq_Bz_r1_quadratic}) become
\begin{eqnarray}
B_r & = & \frac{2 \, a \, r_1^3 \, z}{r^2} \ B_z(r_1,\varphi_0,z_1)
\label{eq_Br_r1_quadratic_spiral} \\
B_z & = & \frac{r_1^2}{r^2} \ B_z(r_1,\varphi_0,z_1) \ ,
\label{eq_Bz_r1_quadratic_spiral}
\end{eqnarray}
and in {\bf model~D},
Eqs.~(\ref{eq_Br_r1_divlinear}) -- (\ref{eq_Bz_r1_divlinear}) become
\begin{eqnarray}
B_r & = & - \frac{n}{n+1} \ \frac{r_1^3}{r^2 \, z} \
\left[ \left( \frac{z}{z_1} \right)^{-n} - \frac{z}{z_1} \right] \
B_z(r_1,\varphi_0,z_1)
\label{eq_Br_r1_divlinear_spiral} \\
B_z & = & \frac{r_1^2}{r^2} \ B_z(r_1,\varphi_0,z_1) \ \cdot
\label{eq_Bz_r1_divlinear_spiral}
\end{eqnarray}
If we assume that $B_z(r_1,\varphi_0,z_1)$ has the same radial variation
as in section~\ref{models_xshape_mfl_r} (see Eqs.~(\ref{eq_B1_r1_even}) and
(\ref{eq_B1_r1_odd})) and that its azimuthal variation is governed by 
Eq.~(\ref{eq_B1_cos}) written as a function of $\varphi_0$, we obtain
\begin{equation}
\label{eq_B1_r1_spiral}
B_z(r_1,\varphi_0,z_1) = B_z(r_1,z_1) \ 
\cos \big( m \, (\varphi_0 - \varphi_\star) \big) \ ,
\end{equation}
with $B_z(r_1,z_1)$ given by Eq.~(\ref{eq_B1_r1_even}) for antisymmetric 
magnetic fields and by Eq.~(\ref{eq_B1_r1_odd}) for symmetric magnetic fields,
$m$ the azimuthal wavenumber and $\varphi_\star$ a free parameter.

\subsection{\label{models_halo_general}Generalization to $z$-dependent pitch angles}

\begin{table*}
\caption{Brief description of our four models of three-dimensional, 
X-shape + spiral magnetic fields in galactic halos.}
\label{table_halo}
\begin{minipage}[t]{2\columnwidth}
\centering
\renewcommand{\footnoterule}{}  
\begin{tabular}{c|cc|c|ccc|ccc}
\hline
\hline
\noalign{\smallskip}
\ \, Model \footnote{
The shape of field lines associated with the poloidal field
is indicated in Table~\ref{table_xshape} (second column),
while the spiraling of field lines is described by Eq.~(\ref{eq_mfl_spiral_rz})
or, more conveniently, by the pitch angle, $p(r,z)$.
} & 
\ \, Reference \footnote{
Prescribed radius or height at which field lines are labeled
by their height or radius, respectively, and by their azimuthal angle.
} &
\ \, Labels of \footnote{
Height or radius of field lines and their azimuthal angle at the reference coordinate.
} &
\ \, Reference \footnote{
Radial or vertical field at the reference coordinate
as a function of field line labels.
} &
$B_r$ & $B_\varphi$ & $B_z$ &
\multicolumn{3}{c}
{\ \quad Free parameters \footnote{
The free parameters of the poloidal field are listed
in Table~\ref{table_xshape} (last column).
Here, we only list the additional free parameters and functions
describing the azimuthal structure.
}\footnote{
$p(r,z)$ is the pitch angle of the magnetic field, 
needed to infer the azimuthal field from the radial field
(see Eq.~\ref{eq_Bphi_alpha_spiral_rz});
$m$ is the azimuthal wavenumber 
and $\varphi_\star(z_1)$ or $\varphi_\star(r_1)$ the fiducial angle 
of the sinusoidal azimuthal modulation at $r_1$ or $z_1$, respectively
(see Eqs.~(\ref{eq_B1_z1_spiral_rz}) and (\ref{eq_B1_r1_spiral_rz})).
Note that $\varphi_\star$ becomes irrelevant in the axisymmetric case
($m = 0$).
}
} \\
& coordinate & field lines & field & & & & 
\multicolumn{3}{c}{and functions} \\
\noalign{\smallskip}
\hline
\noalign{\smallskip}
\ \, A \footnote
{Model~A can only be used in non-axisymmetric magnetic configurations,
with a modification of the azimuthal field component near the $z$-axis,
to avoid a singularity at $r = 0$.
} & 
$r_1 > 0$ & 
$z_1$ (Eq.~\ref{eq_mfl_z1_quadratic}) \ , \ 
$\varphi_1$ (Eq.~\ref{eq_phi1_rz}) &
Eq.~(\ref{eq_B1_z1_spiral_rz}) &
Eq.~(\ref{eq_Br_z1_quadratic_spiral_rz}) & Eq.~(\ref{eq_Bphi_alpha_spiral_rz}) &
Eq.~(\ref{eq_Bz_z1_quadratic_spiral_rz}) &
\ $p(r,z)$ \ & $m$ & $\varphi_\star(z_1)$ \\
\noalign{\smallskip}
\hline
\noalign{\smallskip}
B & $r_1 > 0$ & 
$z_1$ (Eq.~\ref{eq_mfl_z1_divlinear}) \ , \  
$\varphi_1$ (Eq.~\ref{eq_phi1_rz}) &
Eq.~(\ref{eq_B1_z1_spiral_rz}) &
Eq.~(\ref{eq_Br_z1_divlinear_spiral_rz}) & Eq.~(\ref{eq_Bphi_alpha_spiral_rz}) &
Eq.~(\ref{eq_Bz_z1_divlinear_spiral_rz}) &
$p(r,z)$ & $m$ & $\varphi_\star(z_1)$ \\
\noalign{\smallskip}
\hline
\noalign{\smallskip}
C & $z_1 = 0$ & 
$r_1$ (Eq.~\ref{eq_mfl_r1_quadratic}) \ , \  
$\varphi_1$ (Eq.~\ref{eq_phi1_rz}) &
Eq.~(\ref{eq_B1_r1_spiral_rz}) &
Eq.~(\ref{eq_Br_r1_quadratic_spiral_rz}) & Eq.~(\ref{eq_Bphi_alpha_spiral_rz}) &
Eq.~(\ref{eq_Bz_r1_quadratic_spiral_rz}) &
$p(r,z)$ & $m$ & $\varphi_\star(r_1)$ \\
\noalign{\smallskip}
\hline
\noalign{\smallskip}
D & $z_1 = |z_1| \ {\rm sign}~z$ & 
$r_1$ (Eq.~\ref{eq_mfl_r1_divlinear}) \ , \  
$\varphi_1$ (Eq.~\ref{eq_phi1_rz}) &
Eq.~(\ref{eq_B1_r1_spiral_rz}) &
Eq.~(\ref{eq_Br_r1_divlinear_spiral_rz}) & Eq.~(\ref{eq_Bphi_alpha_spiral_rz}) &
Eq.~(\ref{eq_Bz_r1_divlinear_spiral_rz}) &
$p(r,z)$ & $m$ & $\varphi_\star(r_1)$ \\
\noalign{\smallskip}
\hline
\end{tabular}
\end{minipage}
\end{table*}

\begin{figure*}
\hspace*{-1cm}
\includegraphics[scale=0.65]{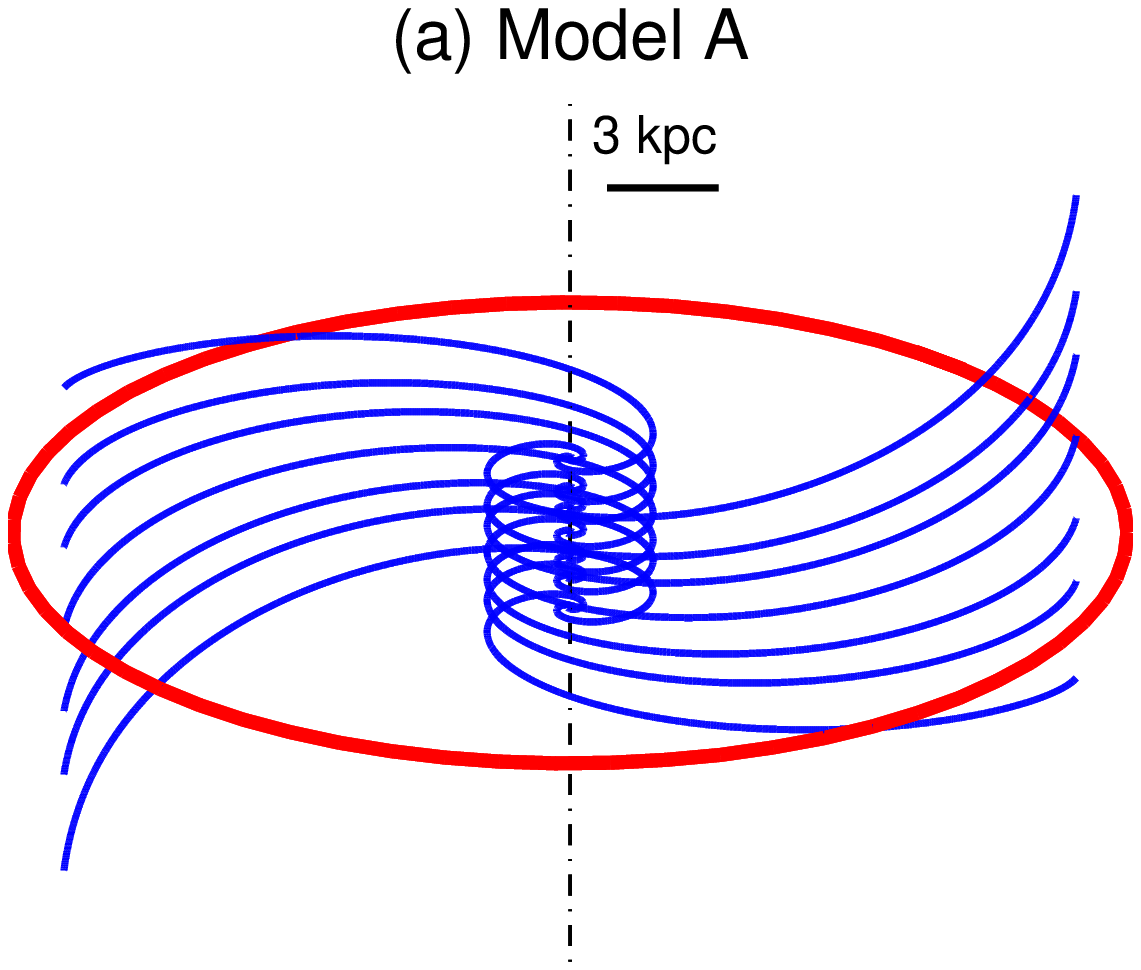}
\hspace*{1cm}
\includegraphics[scale=0.65]{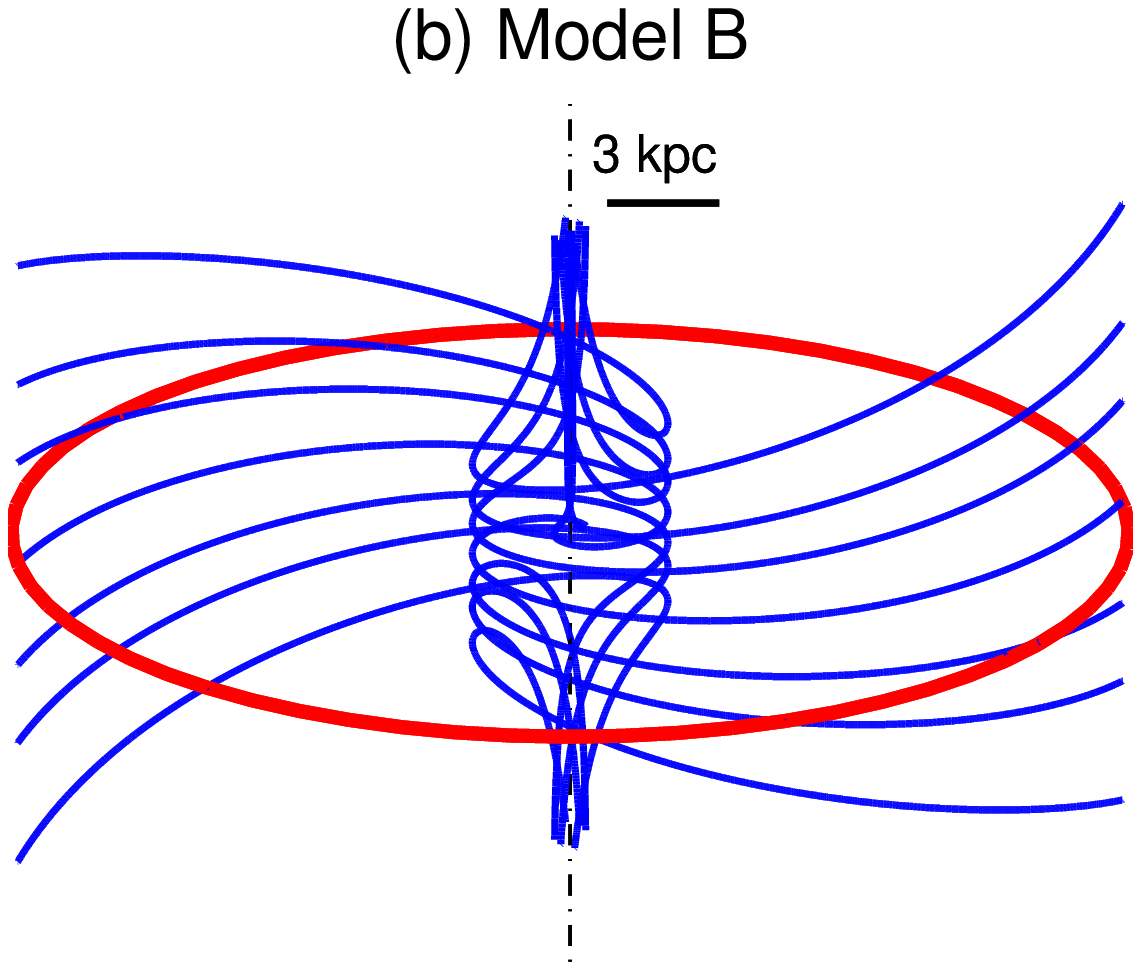}
\\
\hspace*{-1cm}
\includegraphics[scale=0.65]{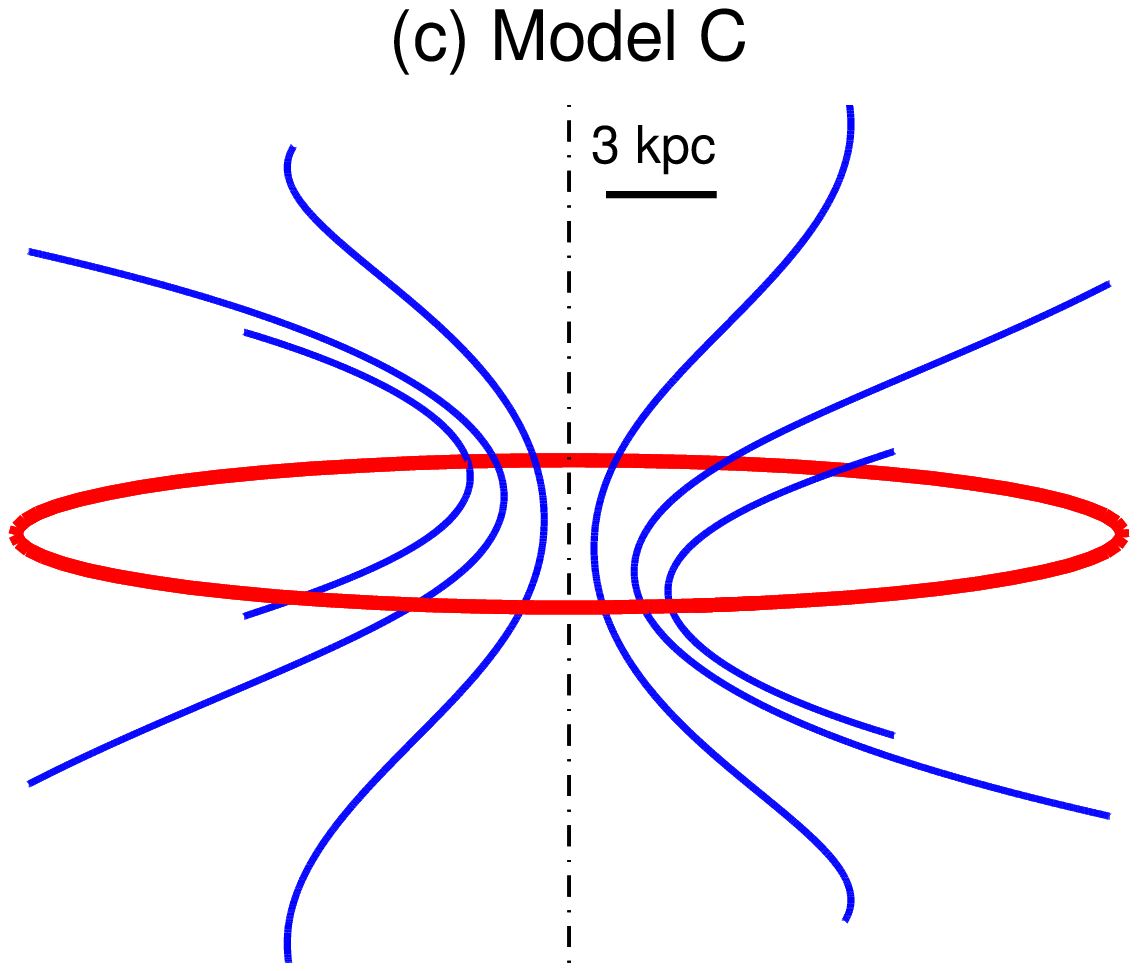}
\hspace*{1cm}
\includegraphics[scale=0.65]{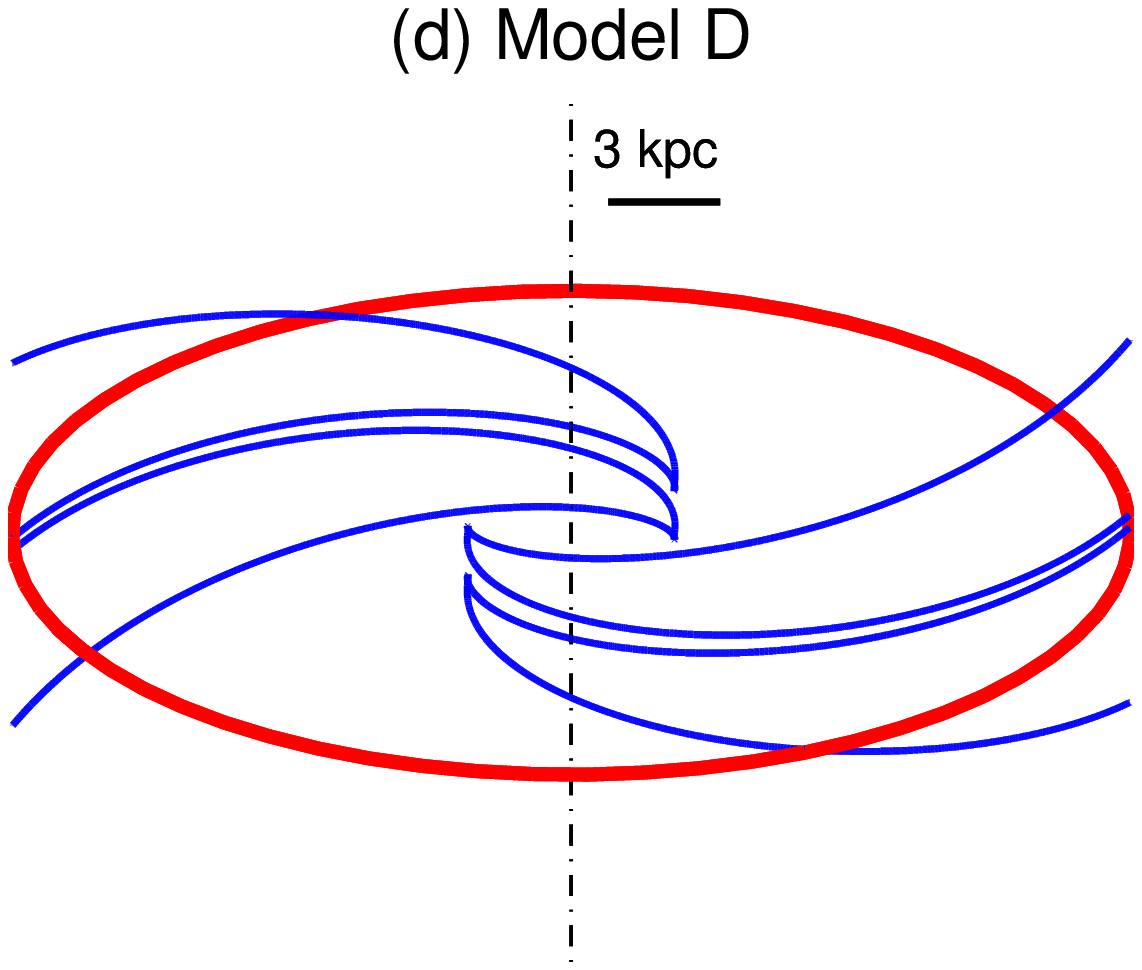}
\caption{Small set of field lines for each of our four models of three-dimensional,
X-shape + spiral magnetic fields in galactic halos,
as seen from an oblique angle.
The parameters of the poloidal field take on the same values 
as in Figure~\ref{figure_xshape},
while the pitch angle is assigned a constant value of $p = -30^\circ$, 
plausibly representative of galactic halos.
The selected field lines correspond to those displayed 
in Figure~\ref{figure_xshape} (i.e., they have the same labels
$(z_1,\varphi_1)$ or $(r_1,\varphi_1)$), except in model~D,
where only four field lines are drawn for clarity.
Each box is a $(30~{\rm kpc})^3$ cube centered on the galactic center.
The trace of the galactic plane is indicated by the red, solid circle,
and the $z$-axis by the vertical, black, dot-dashed line.
}
\label{figure_halo}
\end{figure*}

A remaining limitation of our formalism is that the magnetic pitch angle 
depends only on galactic radius.
Yet, because galactic differential rotation probably decreases 
with distance from the midplane \citep[e.g.,][and references therein]
{levine&hb_08, marasco&f_11, jalocha&bks_11}, 
we expect field lines to become less tightly wound up, 
and hence to have a larger pitch angle, at higher altitudes. 
Note that this is just a qualitative argument, and one should not try 
to directly link the vertical variation of the pitch angle 
to that of the rotational velocity.
For instance, while a vertical decrease in rotational velocity
does indeed reduce the generation of azimuthal field through the shearing 
of radial field,
the vertical velocity gradient itself generates azimuthal field 
through the shearing of vertical field.
Furthermore, the pitch angle depends on other factors, 
such as the intensity of helical turbulence 
and the scale height of the interstellar gas \citep[e.g.,][]{fletcher_10}.

Mathematically, we can allow the pitch angle to vary with both $r$ and $z$
by letting each field line obey an equation similar to Eq.~(\ref{eq_mfl_spiral}),
with $f_\varphi (r)$ replaced by $f_\varphi (r,z)$:
\begin{equation}
\label{eq_mfl_spiral_rz}
\varphi = \varphi_1 + f_\varphi (r,z) \ ,
\end{equation}
where $\varphi_1$ is the azimuthal angle at which the field line passes through
either the reference radius, $r_1$ (in models~A and B),
or the reference height, $z_1$ (in models~C and D).
In both cases, one must have $f_\varphi (r_1,z_1) = 0$.\footnote{
We may no longer take the surface $r = r_0$ to be the normalization surface
on which $f_\varphi$ vanishes,
since not all field lines cross $r_0$ in models~C and D.
This was not an issue in section~\ref{models_halo_spiral},
where we considered, not individual field lines, but the spiral lines
defined by the magnetic field projection onto the galactic plane.
}
The pitch angle, $p(r,z)$, associated with a given winding function, 
$f_\varphi (r,z)$, is easily obtained by differentiation:
\begin{equation}
\label{eq_pitchangle_rz}
\cot p = \frac{r \ d\varphi}{dr}
= r \ \left( \frac{\partial f_\varphi}{\partial r} 
+ \frac{\partial f_\varphi}{\partial z} \ \frac{dz}{dr} \right) \ ,
\end{equation}
where the factor $(dz / dr) = (B_z / B_r)$ depends on the chosen model 
for the poloidal field.
Conversely, the winding function associated with a given pitch angle
is obtained by integrating Eq.~(\ref{eq_pitchangle_rz}) along the poloidal
field line through $(r,z)$:
\begin{equation}
\label{eq_windingfc_rz}
f_\varphi (r,z) = \int_{r_1}^r \ 
\cot p (r',z') \ \frac{dr'}{r'} \ ,
\end{equation}
where $r_1$ is either the reference radius (in models~A and B)
or the radial label of the field line (in models~C and D;
see Eqs.~(\ref{eq_mfl_r1_quadratic}) and (\ref{eq_mfl_r1_divlinear}), 
respectively);
$z_1$ is either the vertical label of the field line (in models~A and B;
see Eqs.~(\ref{eq_mfl_z1_quadratic}) and (\ref{eq_mfl_z1_divlinear}), 
respectively) or the reference height (in models~C and D);
and $z'$ is the height of the field line at radius $r'$
(given by the primed versions of Eqs.~(\ref{eq_mfl_z_quadratic}),
(\ref{eq_mfl_z_divlinear}), (\ref{eq_mfl_r_quadratic}) and
(\ref{eq_mfl_r_divlinear}) for models~A, B, C and D, respectively).\footnote{
In model~A, 
\begin{equation}
z' = z \ \frac{1 + a \, r'^2}{1 + a \, r^2} \ ;
\nonumber
\end{equation}
in model~B,
\begin{equation}
z' = z \ \frac{\displaystyle \left( \frac{r'}{r_1} \right)^{-n} + n \, \frac{r'}{r_1}}
{\displaystyle \left( \frac{r}{r_1} \right)^{-n} + n \, \frac{r}{r_1}} \ ;
\nonumber
\end{equation}
in model~C,
\begin{equation}
z' = {\rm sign}~z \ \sqrt{ 
\frac{1}{a} \ \left[ \frac{r'}{r} \ (1 + a \, z^2) - 1 \right]
} \ ;
\nonumber
\end{equation}
and in model~D, $z'$ is the solution of 
\begin{equation}
\left[ \left( \frac{z'}{z_1} \right)^{-n} + n \, \frac{z'}{z_1} \right] 
= \frac{r'}{r} \ 
\left[ \left( \frac{z}{z_1} \right)^{-n} + n \, \frac{z}{z_1} \right] \ \cdot
\nonumber
\end{equation}
}
A possible simple choice for the pitch angle, which accounts for the decrease 
in differential rotation with height, is
\begin{equation}
\label{eq_pitchangle_z}
p(r,z) = p(z) = p_\infty + (p_0 - p_\infty) \ 
\left( 1+ \left( \frac{|z|}{H_p} \right)^2 \right)^{-1} \ ,
\end{equation}
with $p_0$ the pitch angle at midplane, $p_\infty$ the pitch angle at infinity
and $H_p$ the scale height.

The three components of the magnetic field can now be derived
in the same manner as in section~\ref{models_halo_spiral}.
With
\begin{equation} 
\label{eq_beta_spiral_rz}
\beta = \varphi_1 = \varphi - f_\varphi (r,z)
\end{equation}
chosen for the second Euler potential, Eq.~(\ref{eq_euler}) yields
\begin{eqnarray}
B_r & = & - \frac{1}{r} \ 
\left( \frac{\partial \alpha}{\partial z} \right)_{r,\varphi}
- \frac{1}{r} \ \frac{\partial f_\varphi}{\partial z} \ 
\left( \frac{\partial \alpha}{\partial \varphi} \right)_{r,z} 
\nonumber \\ 
B_\varphi & = & - \frac{\partial f_\varphi}{\partial r} \ 
\left( \frac{\partial \alpha}{\partial z} \right)_{r,\varphi}
+ \frac{\partial f_\varphi}{\partial z} \ 
\left( \frac{\partial \alpha}{\partial r} \right)_{\varphi,z}
\nonumber \\ 
B_z & = & \frac{1}{r} \ 
\left( \frac{\partial \alpha}{\partial r} \right)_{\varphi,z}
+ \frac{1}{r} \ \frac{\partial f_\varphi}{\partial r} \ 
\left( \frac{\partial \alpha}{\partial \varphi} \right)_{r,z} 
\nonumber 
\end{eqnarray}
or, equivalently,
\begin{eqnarray}
B_r & = & - \frac{1}{r} \ 
\left( \frac{\partial \alpha}{\partial z} \right)_{r,\varphi_1}
\label{eq_Br_alpha_spiral_rz} \\
B_\varphi & = & r \ 
\left( \frac{\partial f_\varphi}{\partial r} \ B_r 
+ \frac{\partial f_\varphi}{\partial z} \ B_z \right)
\ = \ \cot p \ B_r
\label{eq_Bphi_alpha_spiral_rz} \\
B_z & = & \frac{1}{r} \ 
\left( \frac{\partial \alpha}{\partial r} \right)_{\varphi_1,z}
\ ,
\label{eq_Bz_alpha_spiral_rz}
\end{eqnarray}
where Eq.~(\ref{eq_pitchangle_rz}) was used to write the second equality 
in Eq.~(\ref{eq_Bphi_alpha_spiral_rz}).
The expressions of $B_r$ and $B_z$ are the same as those obtained 
in section~\ref{models_halo_spiral} (Eqs.~(\ref{eq_Br_alpha_spiral}) 
and (\ref{eq_Bz_alpha_spiral})), with $\varphi_1$ substituting for $\varphi_0$.
The final expression of $B_\varphi$ is also the same as 
in section~\ref{models_halo_spiral} (Eq.~\ref{eq_Bphi_alpha_spiral}),
which is a direct consequence of the definition of the pitch angle.
On the other hand, the intermediate expression of $B_\varphi$ 
is more general, containing not only a term $\propto B_r$ 
(as in Eq.~\ref{eq_Bphi_alpha_spiral}), but also a term $\propto B_z$.
This new term reflects the explicit $z$-dependence of the azimuthal winding 
of field lines, as described by $f_\varphi$ (see Eq.~\ref{eq_mfl_spiral_rz}).
The first equality in Eq.~(\ref{eq_Bphi_alpha_spiral_rz})
can also be viewed as the mathematical statement that the magnetic field 
is tangent to the surfaces defined by Eq.~(\ref{eq_mfl_spiral_rz}).

While Eq.~(\ref{eq_Bphi_alpha_spiral_rz}) provides a general expression
of $B_\varphi$, common to the four different models,
Eqs.~(\ref{eq_Br_alpha_spiral_rz}) and (\ref{eq_Bz_alpha_spiral_rz}) 
need to be specifically worked out for each model.
The resulting expressions of $B_r$ and $B_z$ can be directly taken from
sections~\ref{models_halo_mfl_z} and \ref{models_halo_mfl_r},
with $\varphi_0$ replaced by $\varphi_1$, keeping in mind that
\begin{equation} 
\label{eq_phi1_rz}
\varphi_1 = \varphi - f_\varphi (r,z)
\end{equation}
(see Eq.~\ref{eq_mfl_spiral_rz}).
The final equations are shown below, 
a summary is presented in Table~\ref{table_halo}
and the four models are illustrated in Figure~\ref{figure_halo}.

\subsubsection{\label{models_halo_mfl_z_rz}Poloidal field lines given by $z$ 
as a function of $r$}

As a reminder, $r_1$ is the reference radius,
$z_1$ the vertical label of field lines (dependent on the chosen X-shape model)
and $\varphi_1$ their azimuthal label
(always set by Eq.~(\ref{eq_phi1_rz})).
For a general function $z_1(r,z)$, 
\begin{eqnarray}
B_r & = & \frac{r_1}{r} \ B_r(r_1,\varphi_1,z_1) 
\ \left( \frac{\partial z_1}{\partial z} \right)_r
\label{eq_Br_z1_spiral_rz} \\
B_z & = & - \frac{r_1}{r} \ B_r(r_1,\varphi_1,z_1) 
\ \left( \frac{\partial z_1}{\partial r} \right)_z \ \cdot
\label{eq_Bz_z1_spiral_rz}
\end{eqnarray}
In {\bf model~A}, where $z_1$ is given by Eq.~(\ref{eq_mfl_z1_quadratic}),
\begin{eqnarray}
B_r & = & \frac{r_1}{r} \ \frac{z_1}{z} \ B_r(r_1,\varphi_1,z_1)
\label{eq_Br_z1_quadratic_spiral_rz} \\
B_z & = & \frac{2 \, a \, r_1 \, z_1}{1 + a \, r^2} \ B_r(r_1,\varphi_1,z_1) \ ,
\label{eq_Bz_z1_quadratic_spiral_rz}
\end{eqnarray}
and in {\bf model~B}, where $z_1$ is given by Eq.~(\ref{eq_mfl_z1_divlinear}),
\begin{eqnarray}
B_r & = & \frac{r_1}{r} \ \frac{z_1}{z} \ B_r(r_1,\varphi_1,z_1)
\label{eq_Br_z1_divlinear_spiral_rz} \\
B_z & = & - \frac{n}{n+1} \ \frac{r_1 \, z_1^2}{r^2 \, z} \
\left[ \left( \frac{r}{r_1} \right)^{-n} - \frac{r}{r_1} \right] \
B_r(r_1,\varphi_1,z_1) \ \cdot
\label{eq_Bz_z1_divlinear_spiral_rz}
\end{eqnarray}
In the above equations, 
\begin{equation}
\label{eq_B1_z1_spiral_rz}
B_r(r_1,\varphi_1,z_1) = B_r(r_1,z_1) \ 
\cos \Big( m \, \big( \varphi_1 - \varphi_\star(z_1) \big) \Big) \ ,
\end{equation}
with $B_r(r_1,z_1)$ taken from Eq.~(\ref{eq_B1_z1_even}) for symmetric magnetic fields 
and from Eq.~(\ref{eq_B1_z1_odd}) for antisymmetric magnetic fields,
$m$ the azimuthal wavenumber and $\varphi_\star(z_1)$ a smoothly varying
function of $z_1$.
When $\varphi_\star(z_1)$ reduces to a constant, one recovers exactly 
the results of section~\ref{models_halo_mfl_z}, 
where $\varphi_1 = \varphi_0$.

\subsubsection{\label{models_halo_mfl_r_rz}Poloidal field lines given by $r$ 
as a function of $z$}

Here, $z_1$ is the reference height,
$r_1$ the model-dependent radial label of field lines
and $\varphi_1$ their azimuthal label set by Eq.~(\ref{eq_phi1_rz}).
For a general function $r_1(r,z)$, 
\begin{eqnarray}
B_r & = & - \frac{r_1}{r} \ B_z(r_1,\varphi_1,z_1) 
\ \left( \frac{\partial r_1}{\partial z} \right)_r
\label{eq_Br_r1_spiral_rz} \\
B_z & = & \frac{r_1}{r} \ B_z(r_1,\varphi_1,z_1) 
\ \left( \frac{\partial r_1}{\partial r} \right)_z \ \cdot
\label{eq_Bz_r1_spiral_rz}
\end{eqnarray}
In {\bf model~C}, where $r_1$ is given by Eq.~(\ref{eq_mfl_r1_quadratic}),
\begin{eqnarray}
B_r & = & \frac{2 \, a \, r_1^3 \, z}{r^2} \ B_z(r_1,\varphi_1,z_1)
\label{eq_Br_r1_quadratic_spiral_rz} \\
B_z & = & \frac{r_1^2}{r^2} \ B_z(r_1,\varphi_1,z_1) \ ,
\label{eq_Bz_r1_quadratic_spiral_rz}
\end{eqnarray}
and in {\bf model~D}, where $r_1$ is given by Eq.~(\ref{eq_mfl_r1_divlinear}),
\begin{eqnarray}
B_r & = & - \frac{n}{n+1} \ \frac{r_1^3}{r^2 \, z} \
\left[ \left( \frac{z}{z_1} \right)^{-n} - \frac{z}{z_1} \right] \
B_z(r_1,\varphi_1,z_1)
\label{eq_Br_r1_divlinear_spiral_rz} \\
B_z & = & \frac{r_1^2}{r^2} \ B_z(r_1,\varphi_1,z_1) \ \cdot
\label{eq_Bz_r1_divlinear_spiral_rz}
\end{eqnarray}
In the above equations,
\begin{equation}
\label{eq_B1_r1_spiral_rz}
B_z(r_1,\varphi_1,z_1) = B_z(r_1,z_1) \ 
\cos \Big( m \, \big( \varphi_1 - \varphi_\star(r_1) \big) \Big) \ ,
\end{equation}
with $B_z(r_1,z_1)$ taken from Eq.~(\ref{eq_B1_r1_even}) for antisymmetric 
magnetic fields and from Eq.~(\ref{eq_B1_r1_odd}) for symmetric magnetic fields,
$m$ the azimuthal wavenumber and $\varphi_\star(r_1)$ a smoothly varying
function of $r_1$.
To recover the results of section~\ref{models_halo_mfl_r},
where $\varphi_1 = \varphi_0 + f_\varphi (r_1)$,
it suffices to identify $\varphi_\star(r_1)$ with $f_\varphi (r_1)$ 
to within a constant.

\section{\label{models_disk}Analytical models of three-dimensional magnetic fields
in galactic disks}

Magnetic fields in most galactic disks are approximately horizontal 
and follow a spiral pattern 
\citep[e.g.,][]{wielebinski&k_93, dumke&kwk_95, beck_09, braun&hb_10}.
In section~\ref{models_halo}, we already derived the general expression
of a magnetic field having a spiral horizontal component
(Eqs.~(\ref{eq_Br_alpha_spiral_rz}) -- (\ref{eq_Bz_alpha_spiral_rz})),
and we applied this general expression to our four models of magnetic fields
with an X-shape poloidal component (sections~\ref{models_halo_mfl_z_rz} 
and \ref{models_halo_mfl_r_rz}).
We now use the results of section~\ref{models_halo} to discuss and model 
magnetic fields in galactic disks.

\subsection{\label{models_disk_spiral}Purely horizontal magnetic field 
with a spiral pattern}

In most studies of magnetic fields in galactic disks 
(in particular, in the disk of our own Galaxy), 
the field is assumed to be purely horizontal ($B_z = 0$).
This simplifying assumption has important implications 
for the $r$-dependence of $B_r$, which are generally overlooked.
These implications, which directly emerge from our formalism,
are discussed below.

When $B_z = 0$, Eq.~(\ref{eq_Bz_alpha_spiral_rz}) implies that 
$\alpha(r,\varphi_1,z)$ reduces to a function of $\varphi_1$ and $z$ only,
with the dependence on $r$ dropped out.
It then follows from Eq.~(\ref{eq_Br_alpha_spiral_rz}) that $B_r$ can be written as
\begin{equation}
\label{eq_Br_hor_spiral}
B_r = \frac{1}{r} \ {\rm fc} (\varphi_1,z) \ ,
\end{equation}
or, in the case of a sinusoidal variation with $\varphi_1$,
\begin{equation}
\label{eq_Br_hor_spiral_cos}
B_r = \frac{1}{r} \ {\rm fc} (z) \ 
\cos \Big( m \, \big( \varphi_1 - \varphi_\star(z) \big) \Big) \ ,
\end{equation}
where $\varphi_1$ (given by Eq.~\ref{eq_phi1_rz}) is the azimuthal angle at which 
the field line passing through $(r,\varphi,z)$ crosses a reference radius $r_1$,
$m$ is the azimuthal wavenumber,
and ${\rm fc} (z)$ and $\varphi_\star(z)$ are two function of $z$.
The key point here is that, unless ${\rm fc} (z) = 0$
(trivial case of a perfectly circular field),
$B_r \propto (1/r) \to \infty$ for $r \to 0$, 
regardless of the azimuthal symmetry,
i.e., in axisymmetric ($m = 0$) as well as bisymmetric ($m = 1$) 
and higher-order ($m \ge 2$) configurations.
The physical reason for this singularity is similar to that put forward
in model~A (see below Eq.~\ref{eq_B1_z1_even}), namely, 
all field lines (with all values of $\varphi_1$) in a given horizontal plane 
converge to, or diverge from, the $z$-axis.

Many existing studies completely ignore this $(1/r)$ dependence of $B_r$
\citep[e.g.,][]{han&q_94, sun&rwe_08}.
Others correctly include the $(1/r)$ factor in the expression of $B_r$,
but either they choose to turn a blind eye to the singular behavior 
of $B_r$ at $r = 0$ \citep[e.g.,][]{sofue&f_83, prouza&s_03}
or they artificially get rid of it by replacing, inside a certain radius,
the factor $(1/r)$ by a constant or a regular function of $r$
\citep[e.g.,][]{stanev_97, harari&mr_99, jansson&fwe_09, pshirkov&tkn_11}
-- which leads to a violation of the divergence-free condition.

Like in model~A, there are basically two ways of removing the singularity
at $r = 0$, while keeping the magnetic field divergence-free.
The first way is to let the horizontal field at small radii deviate 
from a pure spiral (defined here by Eq.~\ref{eq_mfl_spiral_rz}),
in such a way that field lines connect two by two away from the $z$-axis
and remain separate from all other field lines.
This re-configuration is possible only for non-axisymmetric fields,
because connecting field lines must necessarily have opposite signs of $B_r$
for the field direction to remain continuous at their connection point.

The alternative approach is to allow for a departure from a strictly 
horizontal magnetic field.
With an adequate vertical component, field lines can bend away 
from the midplane as they approach the $z$-axis
and continue into the halo without ever reaching the $z$-axis.
The advantage of this approach is that it can be implemented in both 
axisymmetric and non-axisymmetric configurations.
Two illustrative examples are provided in section~\ref{models_disk_full}.

\subsection{\label{models_disk_full}Three-dimensional magnetic field 
with a spiral horizontal component}

\begin{table*}
\caption{Brief description of our two models of three-dimensional, 
X-shape + spiral magnetic fields in galactic disks.}
\label{table_disk}
\begin{minipage}[t]{2\columnwidth}
\centering
\renewcommand{\footnoterule}{}  
\begin{tabular}{c|ccc|c|ccc}
\hline
\hline
\noalign{\smallskip}
\ \, Model \footnote{
The free parameters and functions of models~Bd and Dd 
are the same as those of models~B and D, respectively
(see Table~\ref{table_xshape} for the free parameters of the poloidal field
and Table~\ref{table_halo} for the free parameters and functions
describing the azimuthal structure).
The only difference concerns the power-law index of field lines 
asymptotic to the $z$-axis in model~B (now restricted to $n \ge 2$)
or the $r$-axis in model~D (now set to $n = \frac{1}{2}$).
} & 
Shape of &
\ \, Reference \footnote{
Prescribed radius or height at which field lines are labeled
by their height or radius, respectively, and by their azimuthal angle.
} &
\ \, Labels of \footnote{
Height or radius of field lines and their azimuthal angle at the reference coordinate.
} &
\ \, Reference \footnote{
Radial or vertical field at the reference coordinate
as a function of field line labels.
} &
$B_r$ & $B_\varphi$ & $B_z$ \\
& poloidal field lines & coordinate & field lines & field & & & \\
\noalign{\smallskip}
\hline
\noalign{\smallskip}
Bd & Eq.~(\ref{eq_mfl_z_divlinear_disk}) & 
$r_1 > 0$ & 
$z_1$ (Eq.~\ref{eq_mfl_z1_divlinear_disk}) \ , \  
$\varphi_1$ (Eq.~\ref{eq_phi1_rz}) &
Eq.~(\ref{eq_B1_z1_spiral_rz}), with Eq.~(\ref{eq_B1_z1_disk}) &
Eq.~(\ref{eq_Br_z1_divlinear_disk}) & Eq.~(\ref{eq_Bphi_alpha_spiral_rz}) &
Eq.~(\ref{eq_Bz_z1_divlinear_disk}) \\
\noalign{\smallskip}
\hline
\noalign{\smallskip}
Dd & Eq.~(\ref{eq_mfl_r_divlinear}) & 
$z_1 = |z_1| \ {\rm sign}~z$ & 
$r_1$ (Eq.~\ref{eq_mfl_r1_divlinear}) \ , \  
$\varphi_1$ (Eq.~\ref{eq_phi1_rz}) &
Eq.~(\ref{eq_B1_r1_spiral_rz}), with Eq.~(\ref{eq_B1_r1_odd}) &
Eq.~(\ref{eq_Br_r1_divlinear_spiral_rz}) & Eq.~(\ref{eq_Bphi_alpha_spiral_rz}) &
Eq.~(\ref{eq_Bz_r1_divlinear_spiral_rz}) \\
\noalign{\smallskip}
\hline
\end{tabular}
\end{minipage}
\end{table*}

\begin{figure*}
\hspace*{-1cm}
\includegraphics[scale=0.65]{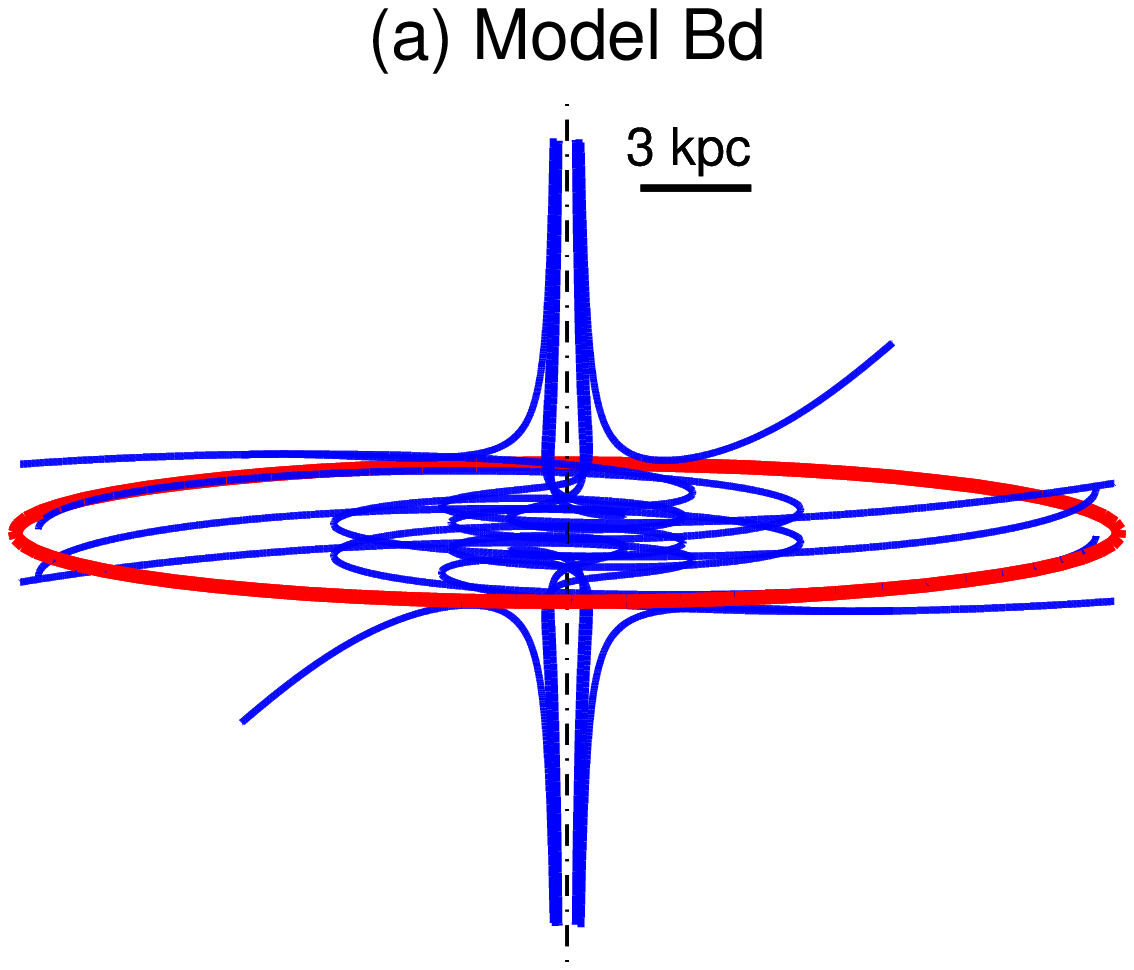}
\hspace*{1cm}
\includegraphics[scale=0.65]{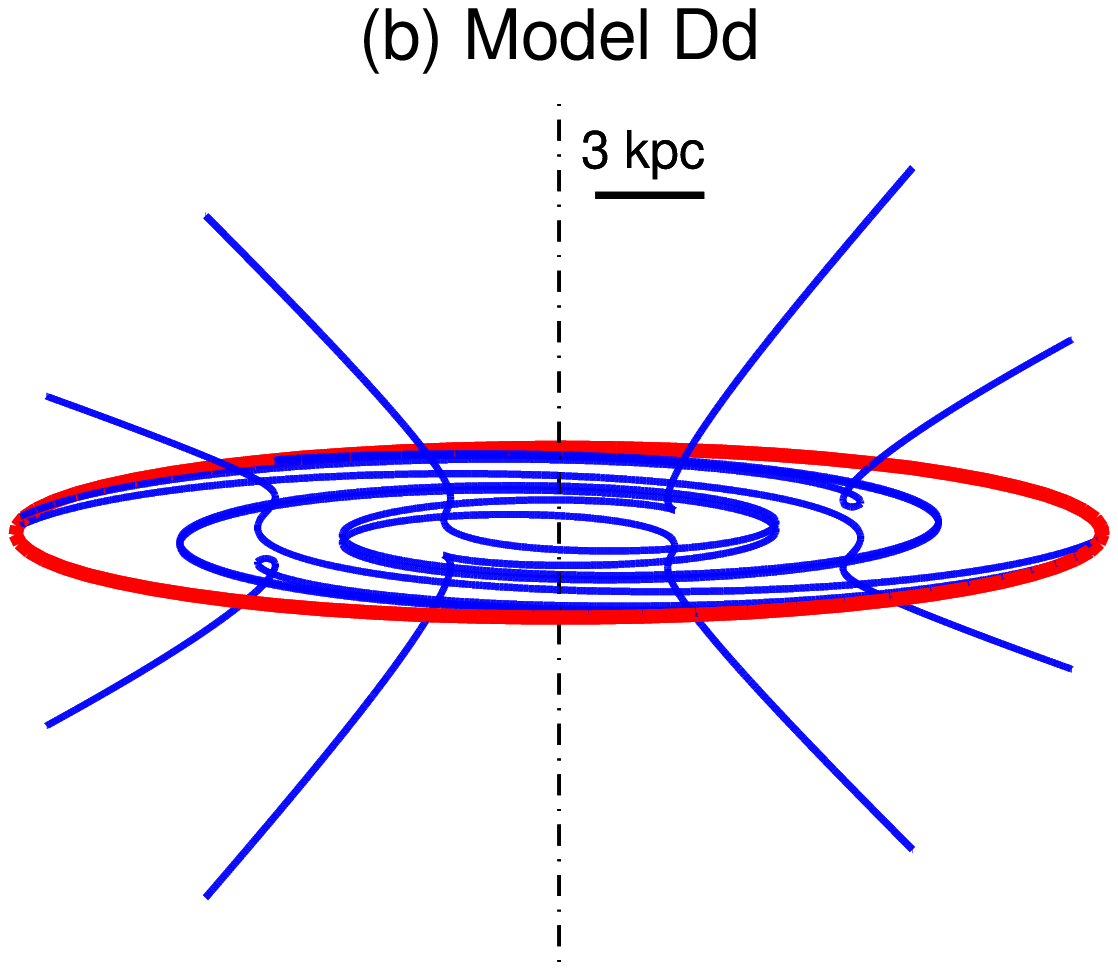}
\caption{Small set of field lines for our two models of three-dimensional,
X-shape + spiral magnetic fields in galactic disks,
as seen from an oblique angle:
(a) model~Bd (poloidal component described by Eq.~\ref{eq_mfl_z_divlinear_disk})
with $r_1 = 3$~kpc and $n = 2$;
and (b) model~Dd (Eq.~\ref{eq_mfl_r_divlinear})
with $|z_1| = 1.5$~kpc and $n = \frac{1}{2}$.
The pitch angle is given by Eq.~(\ref{eq_pitchangle_z})
with $p_0 = -10^\circ$, $p_\infty = -90^\circ$ and $H_p = 3$~kpc.
The selected field lines lie on a given surface of constant 
$\varphi_1$ or $\varphi_1 + \pi$,
and they are separated by a fixed magnetic flux per unit $\varphi_1$;
thus, in model~Bd, their vertical spacing at $r_1$ is determined by 
Eq.~(\ref{eq_B1_z1_disk}) with $H = 1.5$~kpc,
and in model~Dd, their radial spacing at $z_1$ is determined by 
Eq.~(\ref{eq_B1_r1_odd}) with $L = 10$~kpc.
Each box is a $(30~{\rm kpc})^3$ cube centered on the galactic center.
The trace of the galactic plane is indicated by the red, solid circle,
and the $z$-axis by the vertical, black, dot-dashed line.
}
\label{figure_disk}
\end{figure*}

Two of the four models developed in sections~\ref{models_xshape} 
and \ref{models_halo} for magnetic fields in galactic halos 
(models~B and D) have specific properties which should make them 
easily adapted to galactic disks:
unlike model~A, they are regular at $r = 0$;
unlike model~C, they can be symmetric with respect to the midplane;
and they both have nearly horizontal field lines at low $|z|$.
In this section, we present disk versions of these two models, 
which we refer to as models~Bd and Dd, respectively.
The descriptive equations of these two models are listed in Table~\ref{table_disk}, 
and a few representative field lines are plotted in Figure~\ref{figure_disk}.

The notation used for disk fields is the same as that used for halo fields,
namely, $(r_1,\varphi_1,z_1)$ denotes the point where the field line 
passing by $(r,\varphi,z)$ crosses the chosen reference coordinate,
$r_1$ (in model~Bd) or $z_1$ (in model~Dd).
In practice, $z_1$ (in model~Bd) and $r_1$ (in model~Dd) are related to $(r,z)$ 
via the shape of poloidal field lines (see Eqs.~(\ref{eq_mfl_z1_divlinear_disk})
and (\ref{eq_mfl_r1_divlinear}), respectively).
Moreover, $\varphi_1$ is related to $(r,\varphi,z)$ via the winding function 
of the spiral, $f_\varphi (r,z)$ (see Eq.~\ref{eq_phi1_rz}), 
which, in turn, can be expressed in terms of the pitch angle, $p(r,z)$, 
through Eq.~(\ref{eq_windingfc_rz}).
The pitch angle also links the azimuthal field to the radial field
through Eq.~(\ref{eq_Bphi_alpha_spiral_rz}).

{\bf Model~Bd.}
Model~B is not immediately fit to represent disk magnetic fields, 
for two reasons.
First, low-$|z|$ field lines outside the reference radius, $r_1$, 
rise up too steeply into the halo (see Figure~\ref{figure_xshape}b).
Second, the magnetic field strength peaks too high above the midplane 
(the reference field $B_r(r_1,z_1)$ peaks at $|z_1| = H$;
see Eqs.~(\ref{eq_B1_z1_even}) and (\ref{eq_B1_z1_odd})).
As we now show, both shortcomings can easily be overcome.

First, to reduce the slope of low-$|z|$ field lines outside $r_1$,
we replace the linear term $\propto (r/r_1)$ in Eq.~(\ref{eq_mfl_z_divlinear}) 
by a slower-rising term $\propto \sqrt{r/r_1}$.
Under our previous requirement that each field line reaches its minimum height 
at $(r_1,z_1)$, the equation of field lines becomes
\begin{equation}
\label{eq_mfl_z_divlinear_disk}
z = \frac{1}{2n+1} \ z_1 \ 
\left[ \left( \frac{r}{r_1} \right)^{-n} + 2n \, \sqrt{\frac{r}{r_1}} \right] \ ,
\end{equation}
and, after inversion, the vertical label of field lines reads
\begin{equation}
\label{eq_mfl_z1_divlinear_disk}
z_1 = (2n+1) \ z \ 
\left[ \left( \frac{r}{r_1} \right)^{-n} + 2n \, \sqrt{\frac{r}{r_1}} \right]^{-1} 
\ \cdot
\end{equation}
In view of Eqs.~(\ref{eq_Br_z1_spiral_rz}) -- (\ref{eq_Bz_z1_spiral_rz}), 
the poloidal field components are then given by
\begin{eqnarray}
B_r & = & \frac{r_1}{r} \ \frac{z_1}{z} \ B_r(r_1,\varphi_1,z_1)
\label{eq_Br_z1_divlinear_disk} \\
B_z & = & - \frac{n}{2n+1} \ \frac{r_1 \, z_1^2}{r^2 \, z} \
\left[ \left( \frac{r}{r_1} \right)^{-n} - \sqrt{\frac{r}{r_1}} \right] \
B_r(r_1,\varphi_1,z_1) \ \cdot
\label{eq_Bz_z1_divlinear_disk}
\end{eqnarray}
The value of the exponent $n$ in Eq.~(\ref{eq_Bz_z1_divlinear_disk})
is a little more restricted than for halo magnetic fields, 
not because of differences in the expressions of $B_z$ itself
(Eq.~(\ref{eq_Bz_z1_divlinear_disk}) for disk fields
vs. Eq~(\ref{eq_Bz_z1_divlinear_spiral_rz}) for halo fields),
but because of differences in the expressions of $B_r(r_1,z_1)$
(Eq.~(\ref{eq_B1_z1_disk}) for disk fields
vs. Eq.~(\ref{eq_B1_z1_even}) or (\ref{eq_B1_z1_odd}) for halo fields).
Here, the behavior $B_r \propto r^{n-1} \ B_r(r_1,z_1) \propto r^{n-1}$
and $B_z \propto r^{n-2} \ B_r(r_1,z_1) \propto r^{n-2}$
in the limit $r \to 0$ imposes $n \ge 2$.

Second, to bring the peak in magnetic field strength from the halo 
down to the midplane, we keep the expression of $B_r(r_1,\varphi_1,z_1)$
provided by Eq.~(\ref{eq_B1_z1_spiral_rz}),
but we change the vertical profile of the function $B_r(r_1,z_1)$
entering Eq.~(\ref{eq_B1_z1_spiral_rz}) from a linear-exponential 
(as in Eqs.~(\ref{eq_B1_z1_even}) and (\ref{eq_B1_z1_odd})) 
to a pure exponential:
\begin{equation} 
\label{eq_B1_z1_disk}
B_r(r_1,z_1) = B_1 \ \exp \left( - \frac{|z_1|}{H} \right) \ \cdot
\end{equation} 
With this simple choice, $B_r(r_1,z_1)$ is an even function of $z_1$, 
so that the magnetic field is symmetric with respect to the midplane,
as appropriate for galactic disks 
\citep[e.g., ][]{krause_09, braun&hb_10, beck&w_13}.
Let us just mention for completeness that model~Bd could in principle 
be made antisymmetric, simply by mutliplying the right-hand side
of Eq.~(\ref{eq_B1_z1_disk}) by ${\rm sign}~z_1$.

{\bf Model~Dd.}
Model~D is more straightforward to adapt to galactic disks.
Indeed, low-$|z|$ field lines are already closely aligned with the disk plane,
and the magnetic field is clearly much stronger in a narrow band 
around the plane than in the halo (see Figure~\ref{figure_xshape}d).

To obtain a disk version of model~D, it suffices to introduce
two simple restrictions.
First, in the equation of field lines (Eq.~\ref{eq_mfl_r_divlinear}), 
the exponent $n$ must be such that the magnetic field strength peaks, 
while remaining finite, at the midplane. 
Since $B_r \propto z^{2n-1}$ and $B_z \propto z^{2n}$ in the limit $z \to 0$
(see below Eq.~\ref{eq_Bz_r1_divlinear}), 
this can be achieved with, and only with, $n = \frac{1}{2}$.
Second, if disk magnetic fields are indeed symmetric with respect to the midplane,
one must select Eq.~(\ref{eq_B1_r1_odd}) rather than Eq.~(\ref{eq_B1_r1_even}) 
for the radial profile of the reference field $B_z(r_1,z_1)$.

To conclude, we point out a very nice aspect of model~Dd.
Since its field lines are densely packed and quasi-horizontal in the disk,
while they are broadly spread out and X-shaped in the halo,
model~Dd could potentially serve as a single model for complete galactic 
magnetic fields, including both the disk and the halo contributions.
Similarly, although less strikingly, model~Bd could potentially serve 
as a single, complete field model in the cases of galaxies harboring 
strong vertical fields near their centers.
Such a use of models~Dd and Bd would, of course, be conceivable only 
to the extent that the disk and halo fields have the same vertical parity 
(namely, symmetric with respect to the midplane).

\section{\label{polarization_maps}Synchrotron polarization maps}

\begin{figure*}
\hspace*{-1.5cm}
\includegraphics[scale=0.7]{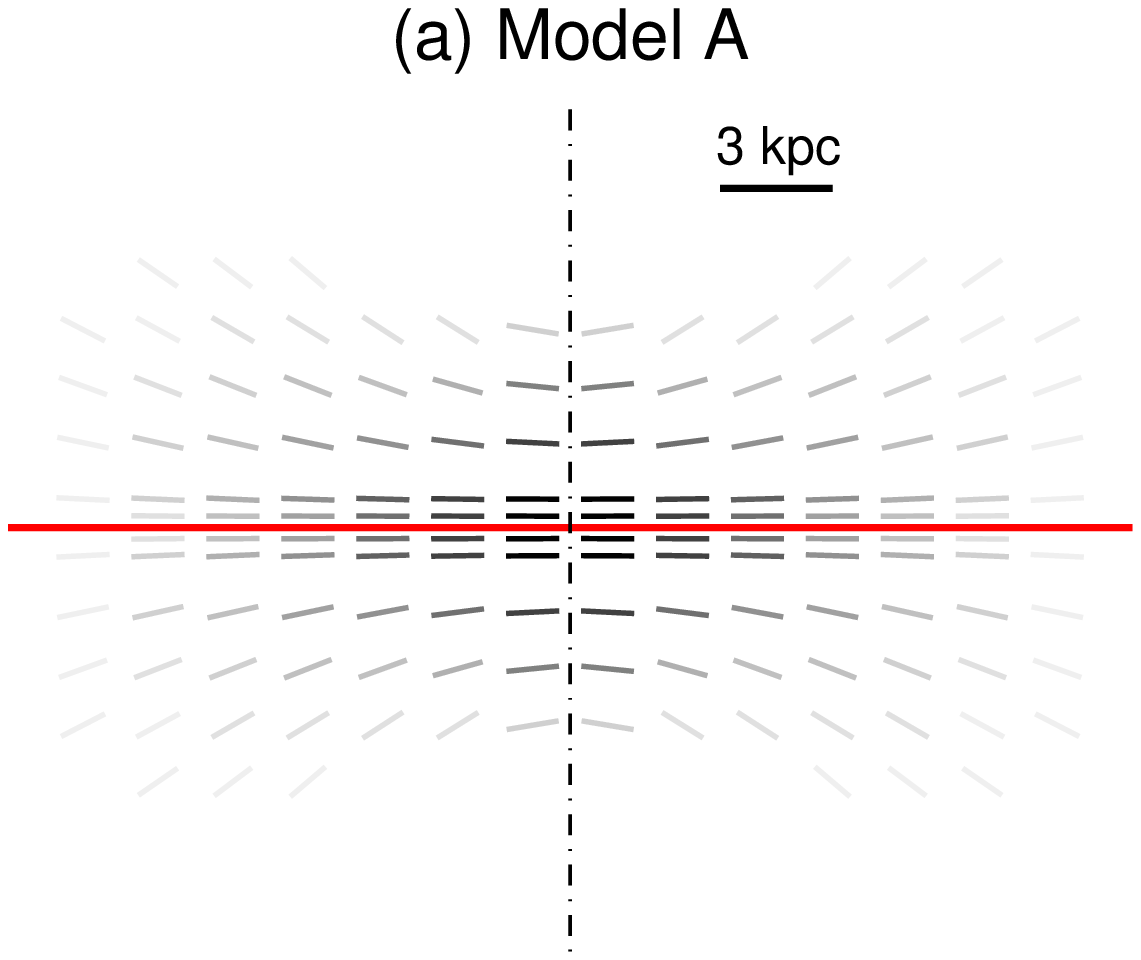}
\includegraphics[scale=0.7]{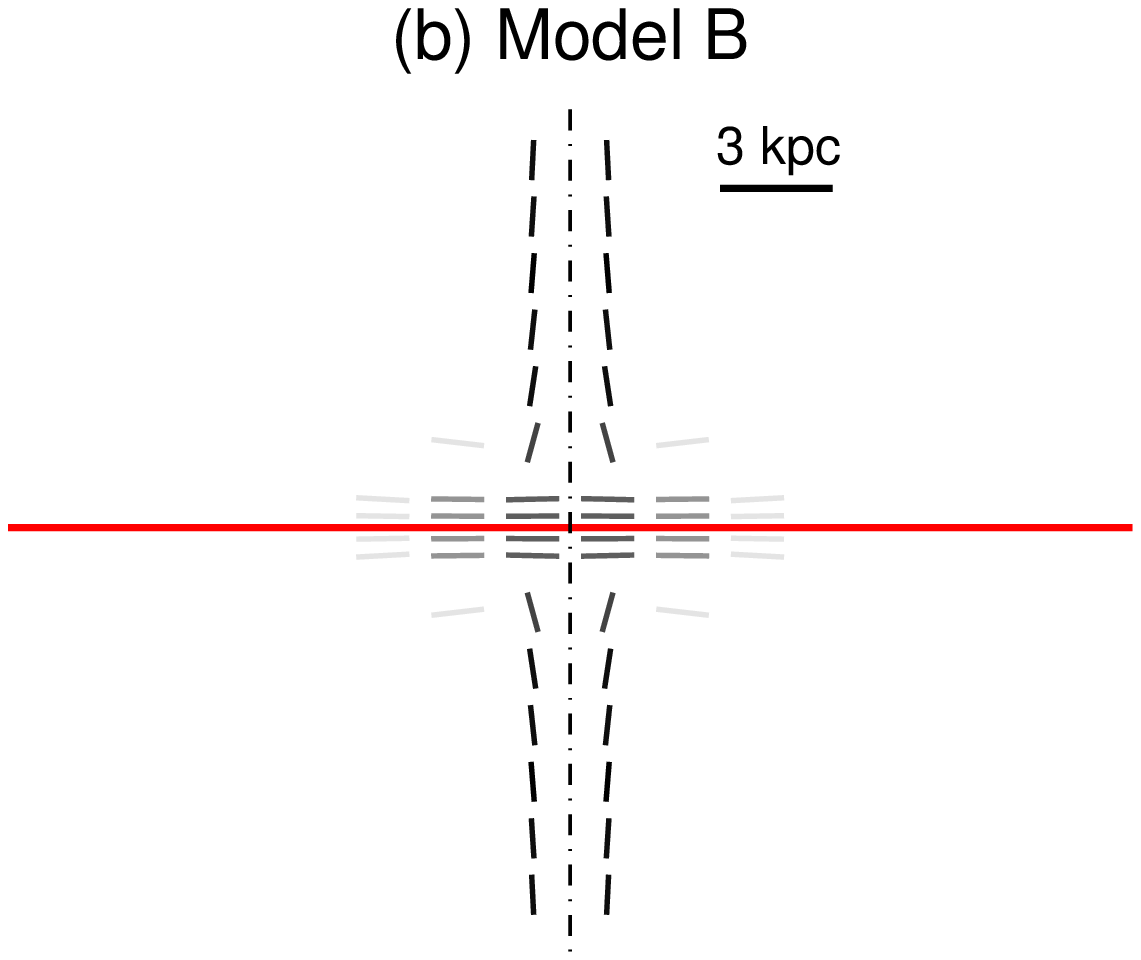}
\bigskip \\
\hspace*{-1.5cm}
\includegraphics[scale=0.7]{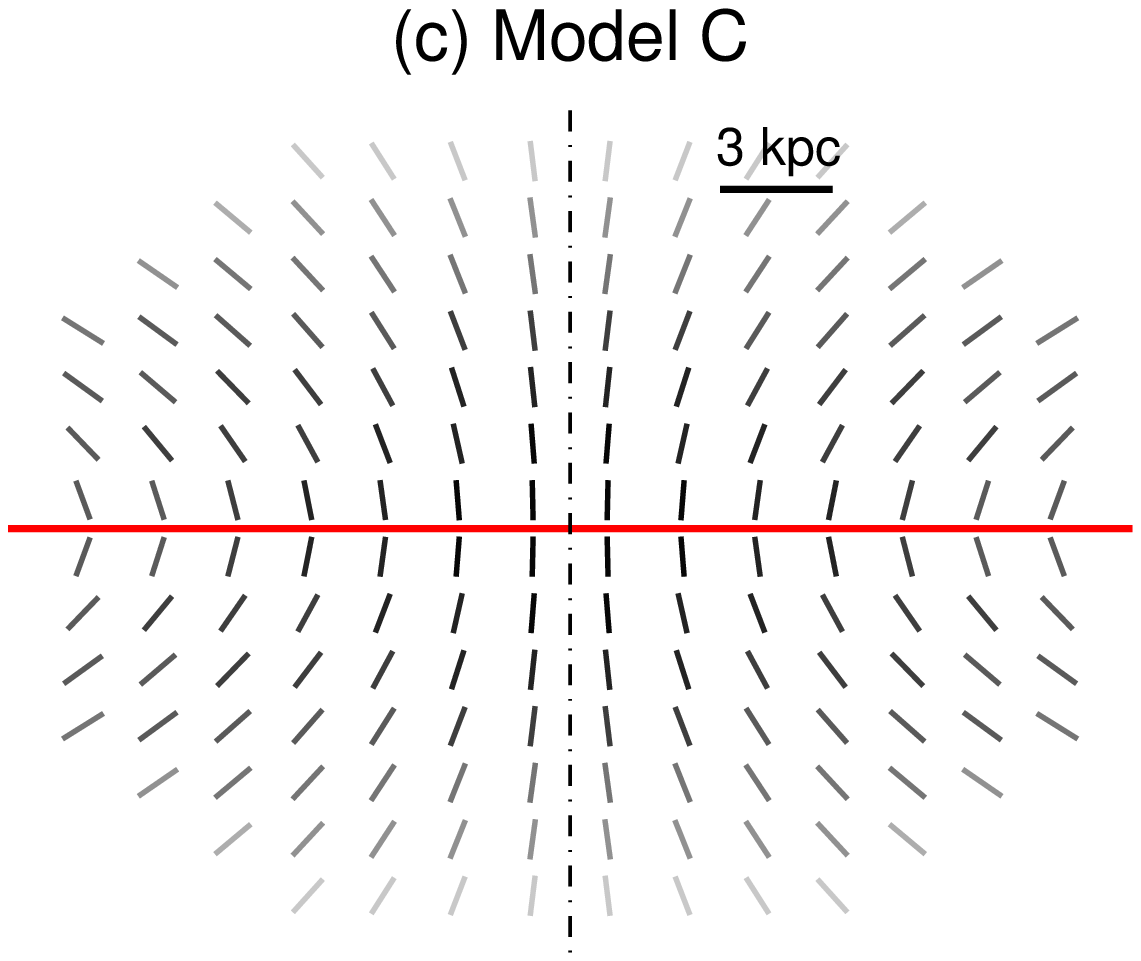}
\includegraphics[scale=0.7]{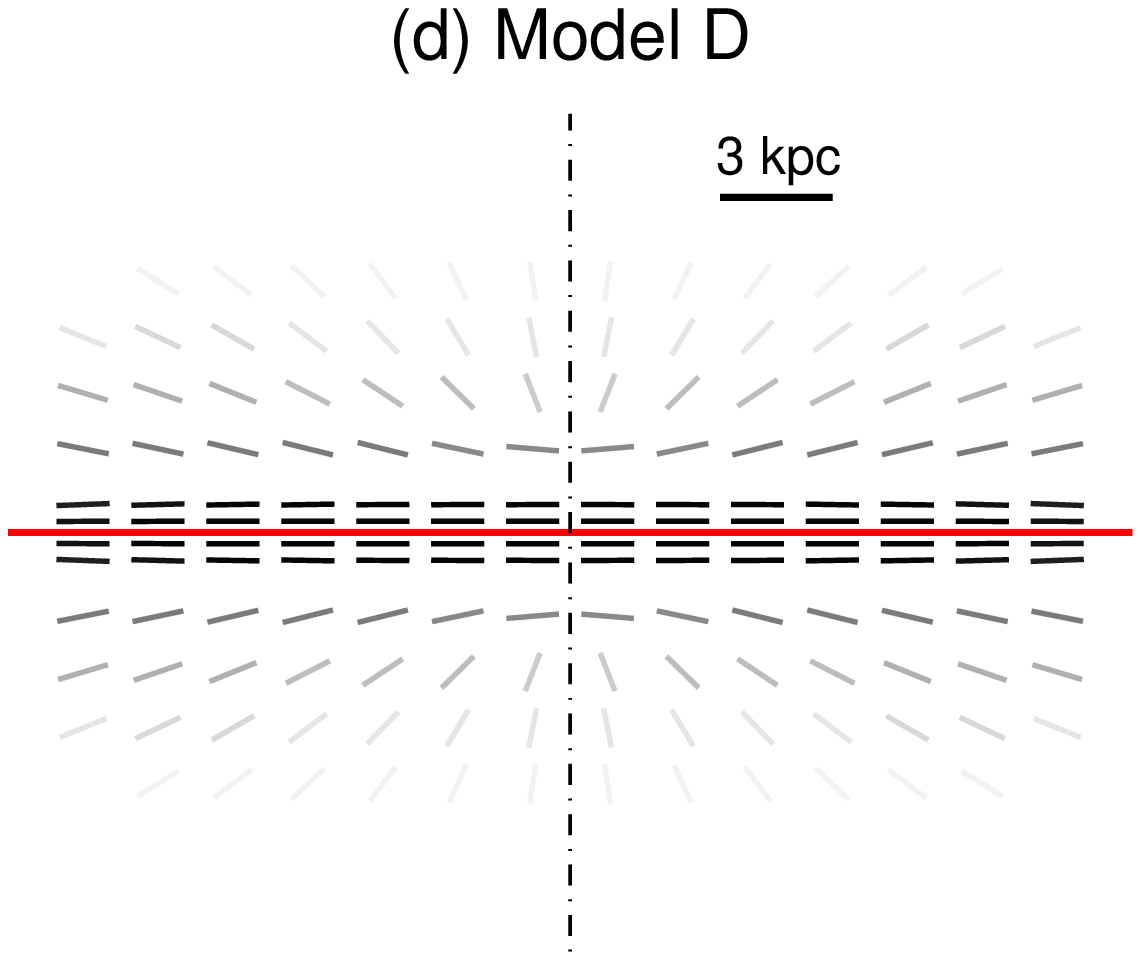}
\caption{
Maps of the normalized synchrotron polarization bars (or headless vectors)
rotated by $90^\circ$ toward a hypothetical external galaxy seen edge-on,
as predicted by our four models of three-dimensional,
X-shape + spiral magnetic fields in galactic halos.
The magnetic field is supposed to be axisymmetric ($m = 0$),
the parameters of its poloidal component have the same values
as in Figures~\ref{figure_xshape} and \ref{figure_halo},
and the pitch angle is given by Eq.~(\ref{eq_pitchangle_z})
with $p_0 = -10^\circ$, $p_\infty = -90^\circ$ and $H_p = 3$~kpc.
The trace of the galactic plane is indicated by the horizontal, red, solid line,
and the $z$-axis by the vertical, black, dot-dashed line.
The rotated polarization bars trace the average magnetic field orientation 
in the plane of the sky, and their level of grey reflects the magnitude
of the polarized intensity (with a logarithmic scale). 
}
\label{figure_polarization}
\end{figure*}

By construction, our models of magnetic fields in galactic halos
are characterized by a poloidal component that has an X shape 
at large $r$ and $|z|$ (see Figure~\ref{figure_xshape}).
However, in the presence of an azimuthal field component,
this X shape can be partly masked by the spiraling of field lines 
(see Figure~\ref{figure_halo}).
How does this affect synchrotron polarization maps?
Does the azimuthal field wash out the X pattern formed 
by the poloidal field alone?

To address this question, we simulate synchrotron polarization maps
of a hypothetical external galaxy seen edge-on,
for our four models in their axisymmetric versions ($m = 0$).
The parameters of the poloidal field are assigned the same values
as in Figures~\ref{figure_xshape} and \ref{figure_halo},
the normalization field, $B_1$, is set to an arbitrary value (irrelevant here),
and the pitch angle, $p$, is assumed to obey Eq.~(\ref{eq_pitchangle_z})
with $p_0 = -10^\circ$, $p_\infty = -90^\circ$ and $H_p = 3$~kpc.
We compute the synchrotron emissivity,
${\cal E} \propto n_e \ B_\perp^{(\gamma + 1)/2}$,
where ${\bf B}_\perp$ is the magnetic field perpendicular 
to the line of sight, 
$n_e$ is the density of relativistic electrons and $\gamma$ is 
the power-law index of the relativistic-electron energy spectrum.
Here, we adopt the conservative value $\gamma = 3$,
and we make the standard double assumption that 
(1) relativistic electrons represent a fixed fraction of 
the cosmic-ray population 
and (2) cosmic rays and magnetic fields are in (energy or pressure) 
equipartition,\footnote{
This double assumption, routinely made by observers, as well as 
the other assumptions underlying our derivation, are not critical 
for the present discussion.
}
so that ${\cal E} \propto B^2 \ B_\perp^2$.
We then compute the associated Stokes parameters $U$ and $Q$,
remembering that synchrotron emission is (partially) linearly polarized
perpendicular to ${\bf B}_\perp$ and assuming that the intrinsic degree 
of linear polarization is uniform throughout the galaxy (consistent with 
the power-law index, $\gamma$, being uniform).
We define a grid of sightlines through the galaxy, integrate the Stokes 
parameters along every sightline and rotate the resulting polarized-intensity 
bar (or headless vector) by $90^\circ$ to recover the magnetic field 
orientation in the plane of the sky.
The polarization maps obtained in this manner are displayed 
in Figure~\ref{figure_polarization}.
We emphasize that these maps do not include any Faraday rotation,
Faraday depolarization or beam depolarization effects.

The synchrotron polarization patterns predicted by models~A, C and D
are reminiscent of the run of poloidal field lines in Figure~\ref{figure_xshape}.
By the same token, they are consistent with the X shapes observed 
in the halos of external edge-on galaxies.
Model~B leads to an X pattern as well, 
but because the polarization bars fade away (logarithmically) 
with decreasing polarized intensity, this X pattern is completely 
overshadowed by the strong polarized emission arising 
from the strong vertical field along the $z$-axis.

The impact of the main model parameters on the polarization maps
can be understood by running a large number of simulations 
with different parameter values.
One of the most critical parameters is the pitch angle, $p$, 
whose effect is found to conform to our physical expectation.
As a general rule, a small pitch angle entails a strong azimuthal field
in regions where the poloidal field (shown in Figure~\ref{figure_xshape})
has a significant radial component (see Eq.~\ref{eq_Bphi_alpha_spiral_rz}).
This strong azimuthal field, in turn, tends to make the magnetic orientation bars 
(i.e., the polarization bars rotated by $90^\circ$)
more horizontal, especially at short projected distances from the $z$-axis, 
where the azimuthal field along most of the line of sight lies close to 
the plane of the sky.
This effect is particularly pronounced toward the central galactic region
in models~B and D, where the magnetic orientation bars are almost perfectly 
horizontal (see Figures~\ref{figure_polarization}b and \ref{figure_polarization}d),
while they would have been nearly vertical for a purely poloidal field.

The scale height of the pitch angle, $H_p$,
is also found to have the expected impact on the polarization maps.
Choosing, somewhat arbitrarily, $H_p = 3$~kpc,
we obtain the X patterns of Figure~\ref{figure_polarization}.
With a larger value of $H_p$, we find that the magnetic orientation bars 
either remain nearly horizontal up to higher altitudes (in models~A, B and D) 
or start turning toward the horizontal closer to the midplane (in model~C).
As a result, the X pattern either emerges higher up (in models~A, B and D) 
or is squeezed toward the midplane (in model~C).

Another parameter affecting the polarization maps is the azimuthal wavenumber, $m$.
The results presented in Figure~\ref{figure_polarization} pertain to 
axisymmetric ($m = 0$) configurations.
Bisymmetric ($m = 1$) configurations are found to lead to a broader range 
of polarization arrangements, in which the X pattern can be either enhanced
or weakened, depending on the viewing angle.
For higher-order ($m \ge 2$) modes, the distribution of magnetic orientation bars
becomes more complex and does not always display a recognizable X pattern.

The variety of polarization structures produced by our four models
may be representative of the variety observed in the real world 
of external galaxies \citep[e.g.,][]{soida_05, krause_09}.
Model~A could {\it a priori} be used to describe the halo fields of galaxies 
like NGC~891, where the X pattern appears to cross the rotation axis 
almost horizontally \citep{krause_09} -- 
under the condition that the magnetic configuration is not axisymmetric.
Model~D would be more appropriate to describe the halo fields of galaxies 
like NGC~4217, where the magnetic orientation bars are nearly horizontal 
near the midplane and gradually turn more vertical with increasing latitude 
away from the central region \citep{soida_05}.
Note, however, that both models~A and D predict a strong polarized intensity 
from the disk, which is not always observed in the polarization maps
of external edge-on galaxies.
This can be because, at low radio frequencies, the (intrinsically polarized)
emission from the disk is subject to strong Faraday depolarization.
Model~C could be the best candidate to describe the halo fields of galaxies
like the starburst galaxy NGC~4631, which appears to host the superposition 
of an X-shape halo field crossing the midplane at nearly right angles
plus a horizontal disk field
(\citeauthor{irwin&bbd_12} \citeyear{irwin&bbd_12}; Mora \& Krause 2013; 
Marita Krause, private communication).
Finally, model~B could probably be used in the case of galaxies 
with strong bipolar outflows from their central regions.
In that respect, it is noteworthy that many, if not all, of the highly 
inclined galaxies studied by \cite{braun&hb_10} show evidence for 
circumnuclear, bipolar-outflow fields.

\section{\label{discussion}Summary}

In this paper, we proposed a general method, based on the Euler potentials,
to derive the analytical expression (as a function of galactocentric 
cylindrical coordinates) of a magnetic field defined by the shape 
of its field lines and by the distribution of its normal component
on a given reference surface (e.g., a vertical cylinder of radius $r_1$ 
or a horizontal plane of height $z_1$).
Utilizing the Euler formalism ensures that the calculated magnetic field 
is divergence-free.

The first and primary application of our method was to magnetic fields 
in galactic halos, which are believed to come with an X shape.
We derived four specific analytical models of X-shape magnetic fields,
starting with purely poloidal fields (section~\ref{models_xshape})
and generalizing to three-dimensional fields that have an X-shape poloidal 
component and a spiral horizontal component (section~\ref{models_halo}).
The pitch angle of the spiral was allowed to vary with both galactic 
radius and height.
The descriptive equations and the free parameters of our four halo models
(models~A, B, C and D) are summarized in Table~\ref{table_halo} 
(see also Table~\ref{table_xshape} for the poloidal component alone) 
and the shapes of their field lines are shown in Figure~\ref{figure_halo} 
(see also Figure~\ref{figure_xshape} for the poloidal component alone).

In models~A and C, field lines have nice and simple X shapes in meridional planes 
(see Figures~\ref{figure_xshape}a and \ref{figure_xshape}c),
and synchrotron polarization maps showcase the best-looking X patterns
(see Figures~\ref{figure_polarization}a and \ref{figure_polarization}c).
Unfortunately, model~A is unphysical, insofar as field lines
from all meridional planes converge to the rotation axis,
thereby causing the radial field component to become infinite.
For non-axisymmetric fields, this difficulty can possibly be circumvented 
with a slight re-organization of field lines at small radii,
but for axisymmetric fields, model~A must definitely be ruled out
(see section~\ref{models_xshape_azim_depend}).
Model~C, for its part, is always physical, but because field lines 
cross the midplane vertically, it is restricted to antisymmetric fields.

In models~B and D, field lines have slightly more complicated shapes
in meridional planes (see Figures~\ref{figure_xshape}b and 
\ref{figure_xshape}d): their outer parts form a well-defined X figure, 
while their inner parts concentrate along the rotation axis (in model~B)
or along the midplane (in model~D).
This geometry is a direct consequence of magnetic flux conservation 
in a setting where field lines tend to converge 
toward the rotation axis (in model~B) or the midplane (in model~D), 
but are not allowed to cross it.
An effect of this concentration of field lines is to overshadow 
the underlying X pattern in synchrotron polarization maps 
(see Figures~\ref{figure_polarization}b and \ref{figure_polarization}d).
Otherwise, the strong vertical field near the rotation axis in model~B 
is not unrealistic, as it could be explained by strong bipolar outflows 
from the central region.
The strong horizontal field near the midplane in model~D is also realistic, 
but it would be more appropriate for a model of the disk field
than a model of the halo field.

A secondary application of our general method was to magnetic fields
in galactic disks, which were assumed to be approximately horizontal 
near the midplane and to have a spiral horizontal component everywhere,
again with spatially varying pitch angle (section~\ref{models_disk}).
Instead of going through an entire new derivation, we took up two of 
our halo models (models~B and D) and modified them slightly in such a way 
as to adapt them to galactic disks.
The descriptive equations of our two disk models (models~Bd and Dd)
are summarized in Table~\ref{table_disk} and the shapes of their field lines 
are shown in Figure~\ref{figure_disk}.

Models~Bd and Dd were both designed to represent disk magnetic fields,
with a high density of nearly horizontal field lines near the midplane.
However, field lines do not remain strictly confined to the disk;
at some point, they leave the disk and enter the halo, where they continue
along X-shaped lines.
Therefore, models~Bd and Dd could potentially represent the disk and halo fields 
combined and, thus, provide complete models of galactic magnetic fields.
One necessary condition for this to be possible would be that the disk 
and halo fields have a common vertical parity (presumably, symmetric 
with respect to the midplane), which would be expected on theoretical grounds,
but does not seem to be fully supported by current observations 
(see section~\ref{models_xshape_vert_sym}).

Anyway, because galactic disks and halos are different structures, 
formed at different times and shaped differently
\citep[e.g.,][]{silk_03, agertz&tm_11, guo&wbd_11, aumer&w_13},
it may be warranted to appeal to separate models for the disk and halo fields.
This keeps open the possibility for these fields to possess different 
vertical parities, azimuthal symmetries, pitch angles...
This also makes it possible to consider the disk or halo field alone,
as required in some specific problems.

Our magnetic field models can serve a variety of purposes,
related to either our Galaxy or external spiral galaxies.
Most obviously, they can be used to test the hypothesis of X-shape 
magnetic fields in galactic halos against existing RM and synchrotron 
observations, and, if relevant, they can help to constrain 
their observational characteristics.
Equally important, they can provide the galactic field models needed 
in a number of theoretical problems and numerical simulations.

The interested readers can resort to one of the specific field models 
derived in this paper (see Table~\ref{table_halo} for the halo fields
and Table~\ref{table_disk} for the disk fields),
but they can also derive their own field models, relying on
Eqs.~(\ref{eq_Br_alpha_spiral_rz}) -- (\ref{eq_Bz_alpha_spiral_rz}),
once they have settled on a given network of field lines.
These new field models may in principle apply to galactic disks or halos, 
come with an X shape or not, have one or several lobes in each quadrant...,
the main difficulty being to find a workable analytical function for 
the first Euler potential, $\alpha$.

\begin{acknowledgement}{}
Katia Ferri\`ere would like to express her gratitude to the many colleagues 
with whom she had lively and enlightening discussions on X-shape magnetic fields:
Robert Braun,
Chris Chy\.zy,
Ralf-J\"urgen Dettmar,
Andrew Fletcher,
Micha{\l} Hanasz,
Marijke Haverkorn,
Anvar Shukurov,
with special thanks to Rainer Beck and Marita Krause.
The authors also thank the referee, Marian Soida, for his careful reading
of the paper and his valuable comments.
\end{acknowledgement}

\bibliographystyle{aa}
\bibliography{BibTex}

\end{document}